\documentclass{aa}

\usepackage{graphicx}
\usepackage{txfonts}
\usepackage{mathrsfs}
\usepackage{subcaption}
\usepackage{longtable}
\usepackage{booktabs}
\usepackage{multirow}
\usepackage{array}
\usepackage{amsmath}
\usepackage[colorlinks,linkcolor=blue,anchorcolor=blue,citecolor=blue]{hyperref}
\usepackage{lscape}
\usepackage{natbib}
\usepackage{color}
\begin{document}

\authorrunning{Wang et al.}
\titlerunning{New Constraints on the Cosmic Star Formation History}

   \title{Constraints on cosmic star formation history via a new modeling of the radio luminosity function of star-forming galaxies}

   \author{Wenjie Wang\inst{1,2},
          Zunli Yuan\inst{1,2}\thanks{Corresponding author \email{yzl@hunnu.edu.cn}}
          \and
          Hongwei Yu\inst{1,2,3}\thanks{Corresponding author \email{hwyu@hunnu.edu.cn}}
          \and
          Jirong Mao\inst{4,5,6}
          }

   \institute{Department of Physics, School of Physics and Electronics, Hunan Normal University, Changsha 410081, China
   \and
     Key Laboratory of Low Dimensional Quantum Structures and Quantum Control, Hunan Normal University, Changsha 410081, China
   \and
    Synergetic Innovation Center for Quantum Effects and Applications, and Institute of Interdisciplinary Studies, Hunan Normal University, Changsha, Hunan 410081, China
    \and
    Yunnan Observatories, Chinese Academy of Sciences, Kunming 650216, China
    \and
    Center for Astronomical Mega-Science, Chinese Academy of Sciences, Beijing 100012, China
    \and
    Key Laboratory for the Structure and Evolution of Celestial Objects, Chinese Academy of Sciences, Kunming  650216, China
    }

   \date{Received; accepted}


  \abstract
   {Radio wavelengths offer a unique possibility to trace the total star-formation rate (SFR) in galaxies, both obscured and unobscured. To probe the dust-unbiased star-formation history, an accurate measurement of the radio luminosity function (LF) for star-forming galaxies (SFGs) is crucial.  }
   {
We make use of an SFG sample (5900 sources) from the Very Large Array (VLA) COSMOS 3 GHz data to perform a new modeling of the radio LF. By integrating the
analytical LF, we aim to calculate the history of the cosmic SFR density (SFRD)  from $z\sim5$ onwards.
   }
   {
For the first time, we use both models of the pure luminosity evolution (PLE) and joint luminosity+density evolution (LADE) to fit the LFs directly to the radio data using a full maximum-likelihood analysis, considering the sample completeness correction. We also incorporate updated observations of local radio LFs and radio source counts into the fitting process to obtain additional constraints.
   }
   {
We find that the PLE model cannot be used to describe the evolution of the radio LF at high redshift ($z>2$). By construct, our LADE models can successfully fit a large amount of data on radio LFs and source counts of SFGs from recent observations. The Akaike information criterion (AIC) also demonstrates that the LADE model is superior to the PLE model. We therefore conclude that density evolution is genuinely indispensable in modeling the evolution of SFG radio LFs. Our SFRD curve shows a good fit to the SFRD points derived by previous radio estimates. In view of the fact that our radio LFs are not biased,  as opposed those of previous studies performed by fitting the $1/V_{\rm max}$ LF points, our SFRD results should be an improvement on these previous estimates. Below $z\sim1.5$, our SFRD matches a published multiwavelength compilation, while our SFRD turns over at a slightly higher redshift ($2<z<2.5$) and falls more rapidly out to high redshift.
   }
   {}


   \maketitle
%

\section{Introduction}

Understanding the formation and evolution of galaxies through cosmic time is a major quest of modern cosmology.
One of the most fundamental processes driving the evolution of galaxies is star formation.
Star formation rate density (SFRD), defined as the amount of stars formed per year in a unit of cosmological volume,
is therefore a critical parameter for  galaxies  \citep[e.g.,][]{lara2010fundamental}.  In recent decades,  a lot of research based on multiwavelength aspects
has been devoted to the accurate measurement of the cosmic evolution of the SFRD.
Remarkable progress has been made, and the SFRD  is now well understood up to $z \sim 3$, when the Universe was no more
than 3 Gyr old \cite[see][for an exhaustive review]{Madau_2014}.
A  consensus has been reached regarding recent history, where the SFRD increases significantly with redshift, reaching a peak at redshift $z \sim 2$,
an epoch known as ``cosmic noon'' \citep[e.g.,][]{2017A&A...602A...5N,2022ApJ...941...10V,2022ApJ...927..204E}.
The picture becomes less clear at higher redshifts. Some studies (mainly based on UV-selected sources) show a steep
decline in the SFRD \citep{2015ApJ...803...34B, 2016MNRAS.459.3812M, 2018ApJ...854...73I}, while studies performed at radio or submillimeter wavelengths
show a flatter SFRD at z > 3 \citep{Gruppioni_2013, RowanRobinson_2016, 2017A&A...602A...5N, 2020A&A...643A...8G}.
Such uncertainty in the evolution of the SFRD  at early cosmic epochs hinders our  understanding of the core mechanism that governs the star formation
rate (SFR) histories of individual galaxies.

The SFR of galaxies can be traced at multiple wavebands, each tracer having its own advantages and disadvantages \citep{kennicutt1998star}.
In dust-free environments, ultraviolet (UV) light originating primarily from young massive stars serves as the most direct tracer of SFR.
UV light can be used to constrain the unobscured star formation out to very high redshifts \citep[e.g.,][]{Mclure_2013, Bowler_2015,
Finkelstein_2015, McLeod_2015, Parsa_2016, Oesch_2018, Ono_2018, Adams_2020, Bowler-2020, Bouwens_2021}.
However, UV observations suffer from dust absorption, which means the SFR measurements  made at these wavelengths are  underestimated
\citep[e.g.,][]{Smail_1997, Bouwens_2009, Riechers_2013, Dudzeviciute_2020}.
When the dust absorbs UV radiation, it gets heated  and reradiates the energy at far-infrared (FIR) wavelengths.
Therefore, FIR emission is ideal for tracing SFR in dust-rich environments \citep[see][]{kennicutt1998star}.
Unfortunately, FIR observations can suffer from poor resolution and source blending.

Deep radio continuum observations are now believed to be very promising tracers; they offer a unique possibility to trace the total SFR in galaxies, both obscured and unobscured.
As such, they may provide the most robust measurement of the
star-formation history of the Universe \citep{2015aska.confE..68J}.
Radio continuum emission, which is not affected by dust obscuration, is also
an end product of the formation of massive stars \citep[e.g.,][]{2022ApJ...941...10V}. After these short-lived \citep[$\tau\leqslant 3\times 10^7$ yr,][]{2021ApJ...914..126M}
stars undergo supernova explosions, the expanding remnants can
accelerate the cosmic ray electrons and give rise to synchrotron radiation at a typical frequency of $ <30$~GHz
\citep[e.g.,][]{Sadler_1989, Condon_1992, Clemens_2008, Tabatabaei_2017}.
Radio emission triggered by the above process  is empirically found to correlate well with the FIR emission of star-forming galaxies (SFGs), known as  the FIR--radio correlation.
This correlation holds over five orders of magnitude in luminosity and extends to high redshifts \citep[][]{Helou_1985, Yun_2001, Bell_2003}, although the redshift evolution is controversial \citep{2010MNRAS.409...92J, Sargent_2010, magnelli2015far, Calistro_2017, Delhaize_2017}.
The FIR--radio correlation can be used to calibrate radio luminosity as a tracer of SFR \citep{Condon_1992}.

In the past several years, deep radio surveys reaching submilli-Jansky(mJy) detection limits have emerged as a powerful tool to investigate the cosmic evolution of SFGs \citep[e.g.,][]{2022ApJ...941...10V, 2022ApJ...927..204E, 2022MNRAS.509.4291M, 2021A&A...656A..48B, 2021MNRAS.500...22B, 2020MNRAS.491.5911O, 2019PASA...36...12U, 2018A&A...620A.192C, 2017MNRAS.469.1912B, 2017A&A...602A...5N, 2009ApJ...690..610S}.
These studies generally measured the radio luminosity functions (LFs) of SFGs; the SFRD can then be estimated by taking the
luminosity weighted integral of the radio LF \citep[e.g.,][]{2022ApJ...941...10V}. As for the form of radio LFs, most of them assumed pure luminosity evolution \citep[PLE; e.g.,][]{smolvcic2009cosmic, 2017A&A...602A...5N, 2020MNRAS.491.5911O, 2022MNRAS.509.4291M}. Very recently, \citet{2022ApJ...941...10V}  combined the COSMOS-XS survey and Very Large Array (VLA)-COSMOS 3 GHz data sets to constrain a radio LF  with
both luminosity and density evolution.  The analytical LFs from these studies  are obtained through fitting the LF points given by the \citet{1968ApJ...151..393S} $1/V_{\rm max}$ estimator. This semi-parametric method has also been adopted by almost all the existing studies \citep[e.g.,][]{2009ApJ...690..610S,2017A&A...602A...5N,2023MNRAS.523.6082C}.
However, given the ordinary precision in the $1/V_{\rm max}$ estimate, the LF points themselves have errors, and fitting to them will propagate the uncertainties to the analytical LFs. In addition, the result would be dependent on the choice of binning in the $1/V_{\rm max}$ method \citep[see][]{fan2001high}.
We believe that a more reliable approach to obtain the analytical LFs is to use a full maximum-likelihood analysis \citep[e.g.,][]{willott2001radio}.

In the present paper, we make use of the VLA-COSMOS 3 GHz data \citep{2017A&A...602A...1S} to measure the radio LFs of SFGs. We use both models of PLE and joint density+luminosity evolution to fit the SFG LFs directly to the radio data using a full maximum-likelihood analysis.
We aim to perform a comprehensive parametric study of the radio LF of SFGs by means of constraints from multiple observational data. Finally, we can probe the dust-unbiased SFRD up to a redshift of $z\sim5$.

The structure of the present paper is outlined below. In Section \ref{data and sample}, we briefly describe the data used. In Section \ref{methods}, we present the method used to constrain the LFs with redshift. In Section \ref{results}, we derive our radio LF evolution through cosmic time and compare it to those in the literature. In Section \ref{Cosmic star formation rate density history}, we calculate the evolution of the cosmic SFRD using the LF models we derived and compare it to the literature. In Section \ref{summary and comclusions}, we summarize our findings and conclusions.

Throughout the paper, we use the flat concordance Lambda cold dark matter ($\Lambda$CDM) cosmology with the following parameters: Hubble constant $H_0 = 70 \rm km s^{-1} Mpc^{-1}$, dark energy density $\Omega_{\Lambda}=0.7$, and matter density $\Omega_{m}=0.3$. We assume the \cite{Chabrier_2003} initial mass function (IMF) to calculate SFRs. We assume a simple power-law radio spectrum for SFGs,  $F_\nu \propto \nu^{-\alpha}$, where $F_\nu$ is the flux density at frequency $\nu$  and $\alpha$ is the spectral index.

\section{Sample}
\label{data and sample}

\begin{figure}
\centering
\includegraphics[width=\columnwidth]{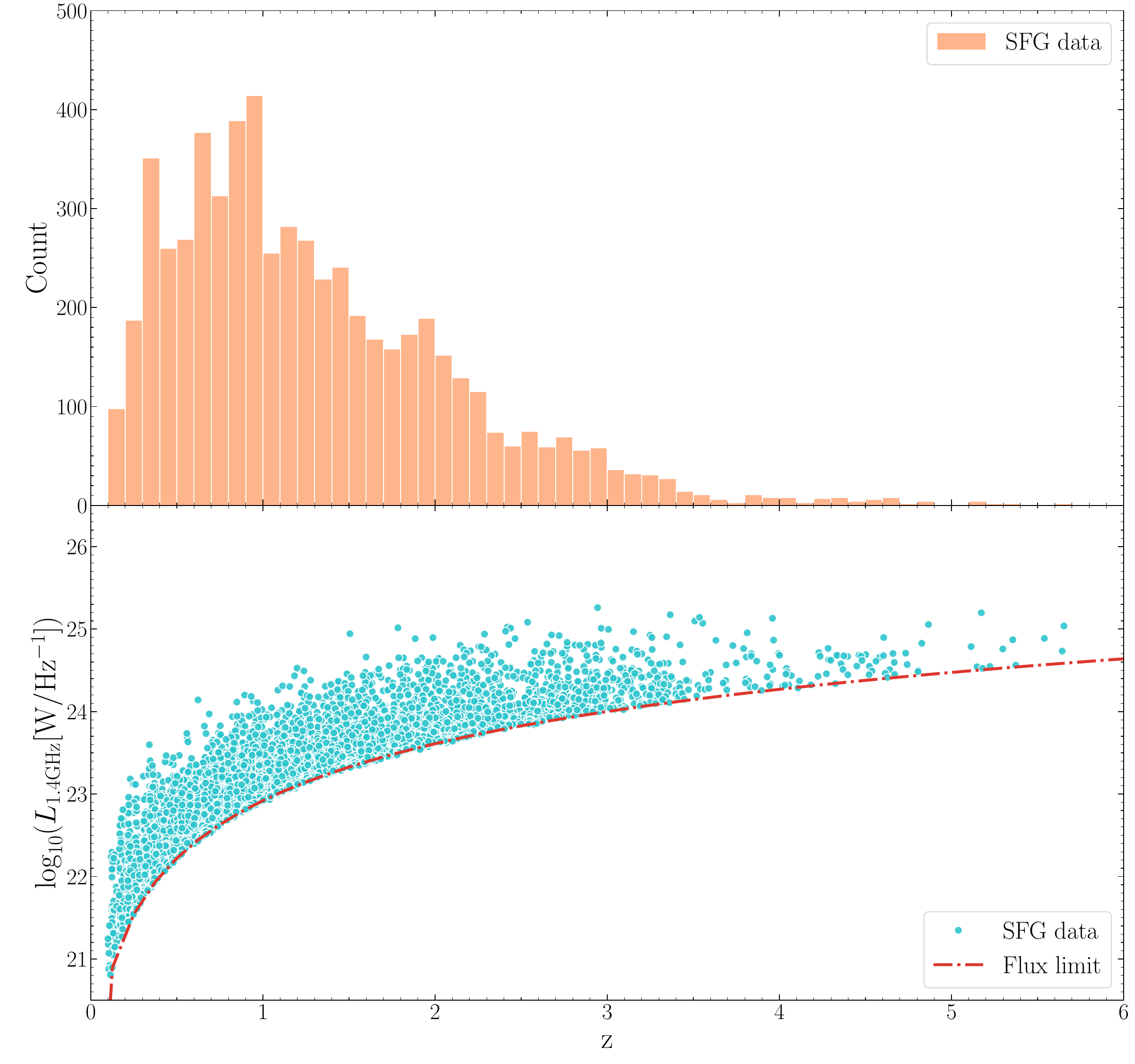}
\caption{
Redshift distribution ($top$) and the scatter plot ($bottom$) of our SFG sample. The red dashed curve indicates the flux limit line $f_{\rm{lim ~ 1.4 GHz}}(z)$.}

\label{fig:Data}
\end{figure}

In our study, we use the same sample of SFGs as presented in \cite{2017A&A...602A...5N}, which was compiled from the continuum data and source catalog release of the VLA-COSMOS 3 GHz Large Project survey \citep{2017A&A...602A...1S}. The sample of SFGs was selected via radio emission and complemented with ancillary data from the comprehensive multiwavelength coverage of COSMOS. The data analysis and multiband association procedure are fully described in \cite{2017A&A...602A...5N}, and we refer readers to that publication for a complete description. Here we summarize some key points about the sample.

The VLA-COSMOS 3 GHz Large Project survey utilized 384 hours of VLA A+C array observations in the S band to obtain radio data. The survey covered a uniform rms noise of 2.3 $\mu$Jy $\text{beam}^{-1}$ and had an angular resolution of $0^{\prime \prime}_{\cdot} 75$ across the 2 square degrees of COSMOS. The final catalog contains 10 830 radio sources.
Taking into account the fraction of spurious sources, an $11\%$ incompleteness in the counterpart sample is estimated. Therefore, a total of 7729 radio sources with assigned COSMOS 2015 counterparts were used. About 35\% of these radio sources have spectroscopic redshifts, and photometric redshifts were used for the remainder of the sample. According to \cite{Delvecchio_2017}, sources were classified as radio-excess if the value of $r$ deviates by more than $3\sigma$ from the peak of the distribution obtained as a function of redshift, that is,
\begin{equation}
r=\log \left(\frac{L_{1.4 \mathrm{GHz}}\left[\mathrm{W} \mathrm{Hz}^{-1}\right]}{SFR_{\mathrm{IR}}\left[M_{\odot} \mathrm{yr}^{-1}\right]}\right)>22 \times(1+z)^{0.013}
\label{eq:excess_cut}
.\end{equation}
According to this criterion, they were able to distinguish 1814 sources (23\%) that are primarily emitting due to AGN activity in the radio.
The sample consists of 5915 SFGs that do not exhibit radio excess. All the sources have a (spectroscopic or photometric) redshift
and the rest-frame 1.4 GHz luminosity.
The redshift distribution of the SFG sample as well as its scatter plot
are shown in Figure \ref{fig:Data}. The red dashed curve indicates the 1.4 GHz flux limit line defined as
\begin{eqnarray}
        \label{flim}
        f_{\rm{lim ~ 1.4 GHz}}(z)=\frac{4 \pi D_{L}^{2}}{(1+z)^{1-\alpha}}\left(\frac{3 ~ \mathrm{GHz}}{1.4 ~ \mathrm{GHz}}\right)^{\alpha}F_{\rm{lim ~ 3GHz}},
\end{eqnarray}
where $D_L$ represents the luminosity distance at redshift $z$, $F_{\rm{lim ~ 3GHz}} = 11.5\mu\rm{Jy} $ is the 5$\sigma$ detection limit of the survey at 3 GHz, and the spectral index $\alpha$ is set to $0.7$. We have excluded all sources below the flux limit line, and the total number of sources used in this work is 5900.

\section{Methods}
\label{methods}

\subsection{Luminosity function and likelihood function}

The LF $\Phi(z, L)$ is a measurement  of the number of sources per unit comoving volume  per unit logarithmic luminosity interval:
\begin{eqnarray}
\label{LFdf}
\Phi(z,L)=\frac{d^{2}N}{dVd\log_{10}L}.
\end{eqnarray}
Given an analytical form with parameters $\mathbf{\theta}$ for the LF, $\Phi(z, L|\mathbf{\theta})$, the maximum-likelihood solution to $\mathbf{\theta}$  is
obtained by minimizing the negative logarithmic likelihood function $S$. Following \cite{marshall1983analysis} and \citep{fan2001high}, $S$ can be written as
\begin{eqnarray}
\label{likelihood1}
\begin{aligned}
S=&-2\!\!\sum_{i}^{n}\!\ln[\Phi(z_{i},L_{i})p(z_{i},L_{i})] \\
&+2\!\!\!\int\!\!\!\!\int_{W}\!\!\!\Phi(z,L)p(z,L)\Omega\frac{dV}{dz}dzdL,
\end{aligned}
\end{eqnarray}
where $p(z,L)$ is the selection probability of the SFG as a function of redshift and luminosity, and $W$ is the survey region. The inclusion of the selection probability in equation (\ref{likelihood1}) accounts for
the fact that the sample is incomplete near the flux limit. The symbol $\Omega$ represents the solid angle covered by the survey, and $dV/dz$ denotes the differential comoving volume per unit solid angle, as defined by \cite{hogg1999distance}.

For our SFG sample, $p(z,L)$ can be estimated by
\begin{equation}
p(z,L)= C_{\text{radio}}[F_{3~\text{GHz}}(z)] \times C_{\text{opt}}(z),
\end{equation}
where $C_{\text{radio}}$  is the completeness of the VLA-COSMOS 3 GHz  radio catalog as a function of the flux density $F_{3~\text{GHz}}$,
and $C_{\text{opt}}$ is the completeness owing to radio sources without assigned optical-NIR counterparts \citep{2017A&A...602A...5N}.
We adopt the calculations of $C_{\text{radio}}$ and $C_{\text{opt}}$ given by \cite{2017A&A...602A...5N}, and refer the interested reader to their Fig. 2 for more details.

To estimate the integration term in Equation (\ref{likelihood1}), one needs to  find the function values for $p(z,L)$ at given pairs of $(z,L)$. We achieve this using an interpolation method. Firstly, we set a two-dimensional (2D) grid of $50 \times 50$ in the $\log L-z$ space. For each grid point ($\log L_i, z_i$), we can derive its flux density $F_i$ from $L_i$ by assuming $\alpha=0.7$. We can then estimate the corresponding $C_{\text{radio}}$ and $C_{\text{opt}}$ through a one-dimensional linear interpolation method using the observed value from \cite{2017A&A...602A...5N}.
Finally, we have the values for $p(z,L)$ at the $50 \times 50$ grid points, which are used to perform the 2D linear interpolation to estimate the function value of $p(z,L)$.

Following the method of \cite{willott2001radio} and \cite{yuan2017mixture}, we incorporate the most recent observations of the local radio LFs and source counts (see section \ref{SC}) into the fitting process to obtain additional constraints.
The local radio LF (LRLF) and the source counts (SCs) are one-dimensional functions, and their $\chi^{2}$ value is calculated as
\begin{eqnarray}
\label{chi2}
\begin{aligned}
\chi^{2}=\sum_{i=1}^{n}\left(\frac{f_{\text {data } i}-f_{\bmod i}}{\sigma_{\text {data } i}}\right)^{2},
\end{aligned}
\end{eqnarray}
where $f_{\text {data } i}$  represents the value of the data in the $i$th bin, and $f_{\text {mod } i}$  and $\sigma_{\text {data } i}$  are  the model value and  data error in the ith bin,
respectively. As $\chi^2$  is related to a likelihood by $\chi^2 = -2 \ln(\text{likelihood})$ \citep[i.e., the same form as $S$;][]{willott2001radio}, we can
 define a new function $S_{\text{all}}$, which combines the constraints from all three types of data (i.e., the SFG sample, LRLF, and SC data). The expression is as follows
\begin{eqnarray}
\label{chi2all}
S_{\text{all}}=S + \chi^2_{\text{LRLF}} + A_0\chi^2_{\text{SC}},
\end{eqnarray}
where $\chi^2_{\text{LRLF}}$ and $\chi^2_{\mathrm{SC}}$ denote the value of $\chi^2$ for the local radio LFs and source counts, respectively. Because we use  three different types of data to estimate $S_{\mathrm{all}}$, we need to balance the statistical weight for each term in Equation (\ref{chi2all}).
We chose an $A_0$ so that the value of $A_0\chi^2_{\mathrm{SC}}$  is approximately equal to that of $\chi^2_{\text{LRLF}}$. This yields values of about 10-40 for our calculations. We find that varying $A_0$ does not significantly bias our final results \citep[also see][]{1996ApJ...473..595K}.
Using Equation (\ref{chi2all}), we can obtain the best-fit parameters for LFs by numerically minimizing the objective function $S_{\text{all}}$. Here we  adopt a Bayesian method as in our previous papers \citep[e.g.,][]{yuan2016mixture}. This latter enables us to determine the best estimates for the model parameters and their probability distribution \citep[also see][]{lewis2002cosmological, yuan2017mixture}. We use uniform (so-called ``uninformative'') priors on the parameters, and employ the MCMC sampling algorithm available in the Python package {\sc{emcee}} \citep{foreman2013emcee} to estimate the best-fit parameters.

\begin{figure}
\centering
\includegraphics[width=\columnwidth]{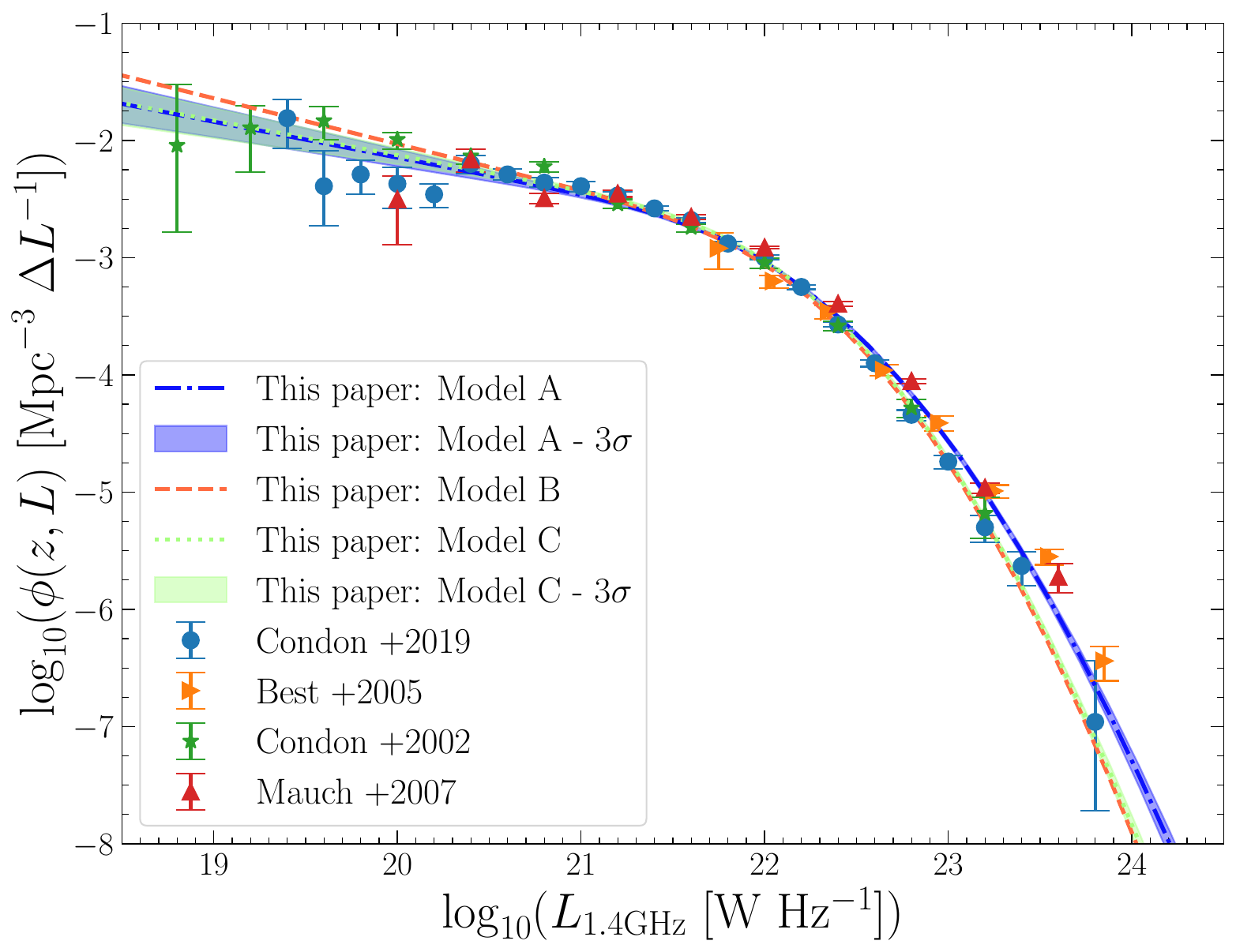}
\caption{
Local radio LF at 1.4 GHz of SFGs from several surveys with different observed areas and sensitivities (colored data points). The colored lines show the fits to the combined data from our models.}
\label{fig:LocalLuminosityFunction_LF}
\end{figure}

\subsection{Local luminosity functions and radio source counts}
\label{SC}
The local LFs at 1.4 GHz have been well determined for SFGs thanks to the combined use of large radio surveys, such as NVSS (NRAO VLA Sky Survey) and FIRST (Faint Images of the Radio Sky at Twenty centimeters), and large-area
spectroscopic surveys. In the present work, we simultaneously use the local SFG LFs from \citet{Condon_2002}, \citet{best2005sample}, \citet{Mauch_2007}, and \citet{Condon_2019} (see Figure \ref{fig:LocalLuminosityFunction_LF}) to calculate $\chi^2_{\text{LRLF}}$ in Equation (\ref{chi2all}).

In addition to local LFs, the observed radio source counts can provide an important constraint to the modeling of SFG LFs.
In the past several years, deep radio surveys have emerged that reach submJy detection limits, enabling investigation of the faint source counts  \citep[e.g.,][]{2017A&A...602A...2S,2020MNRAS.491.1127O,2021A&A...648A...5M, 2021ApJ...907....5V,2021ApJ...909..193M,2023MNRAS.520.2668H}.
The source counts, denoted $n(F_\nu)$, represent the number of sources per flux density ($F_\nu$) per steradian. The shape of $n(F_\nu)$ is closely related to the evolutionary properties of the source as well as the geometry of the Universe \citep{Padovani_2016}. Typically, the counts are Euclidean normalized by multiplying by $F^{2.5}_\nu$ \citep[e.g.,][]{2010A&ARv..18....1D}. According to \cite{Padovani_2016} and \cite{yuan2017mixture}, we can relate the source counts of SFGs to their LF using the following equation:
\begin{eqnarray}
\label{sc}
\begin{aligned}
\frac{n(F_\nu)}{4\pi} = 4\pi\frac c{H_0}  \int_{z_\text{min}(F_\nu)}^{z_\text{max}(F_\nu)} \frac{\Phi(z,L(F_\nu,z))D_L^4(z)dz}{(1+z)^{(3-\alpha)}\sqrt{\Omega_\mathrm{m}(1+z)^3+\Omega_\Lambda}},
\end{aligned}
\end{eqnarray}
where $c$ is the speed of light, $\Phi(z,L)$ is the LF, $D_L(z)$ is the luminosity distance, $z_\text{min}$ and $z_\text{max}$ represent the range of integration in redshift, and $\alpha$ is the spectral index.

In this work, we use the observed source counts from \citet{2020ApJ...903..139A} and \cite{2023MNRAS.520.2668H} to provide an additional constraint in our analysis.
The \citet{2020ApJ...903..139A} $3$ GHz source counts are measured based on the ultrafaint (reaching a $5\sigma$ flux limit of $\sim$2.7 $\mu$Jy beam$^{-1}$ within the center of the $3$ GHz image) radio population detected in the Karl G. Jansky Very Large Array COSMOS-XS survey. The \cite{2023MNRAS.520.2668H} $1.4$ GHz source counts are measured based on the continuum early science data release of the MeerKAT International Gigahertz Tiered Extragalactic Exploration (MIGHTEE) survey in the COSMOS and XMM-LSS fields. The MIGHTEE sources
were divided into three subsets: SFGs, AGNs, and unclassified sources.
\cite{2023MNRAS.520.2668H} considered two cases: (1) the unclassified sources are assumed to be a mix of SFGs and AGNs based on the flux density ratio of classified sources, and (2) the unclassified sources are regarded as SFGs. The source counts for the two cases are presented in Table 1 (for the COSMOS field) and Table 2 (for the XMM-LSS field) presented by these latter authors, respectively. In this work, we use the first case, where the
unclassified or unmatched sources are assumed to have the same split
between SFGs and AGN as the classified sources at the given flux density.
The source counts are shown in the $\mathrm{SC}_{\mathrm{SFG,~ratio}}$ column in Table 1 and Table 2 of \cite{2023MNRAS.520.2668H}.
For the convenience of calculation, the above source counts are unified to $1.4$ GHz by assuming a spectral index of 0.7.

\subsection{Models for the luminosity function of star-forming galaxies}
\label{Form}

Without loss of generality, the SFG LF can be written as
\begin{eqnarray}
\label{aaa}
\Phi(z,L)=e_1(z)\phi(z=0,L/e_2(z),\eta^j),
\end{eqnarray}
where $e_1(z)$ and $e_2(z)$ denote the density evolution (DE) and luminosity evolution (LE) functions of redshift, respectively, and $\eta^j$ represents the parameters that determine the shape of the LF. If the values of $\eta^j$ are constant, this indicates that the shape of the radio LF is unchanged with redshift. Conversely, if $\eta^j$ exhibits a redshift dependence, this implies luminosity-dependent density evolution \citep[see][for a more detailed discussion]{singal2013radio,singal2014gamma}. We assume the shape of the LF to remain unchanged (i.e., $\eta^j$ is constant) as in many other studies \citep[e.g.,][]{2017A&A...602A...5N,2022ApJ...941...10V}.

Following previous work \citep[e.g.,][]{smolvcic2009cosmic,Gruppioni_2013,2022ApJ...941...10V}, the SFG local LF $\phi(z=0,L/e_2(z=0))$  is described by a modified-Schechter function from \citet{saunders199060}:
\begin{eqnarray}
\label{rho}
\begin{aligned}
\phi(z&=0,L/e_2(z=0))=\frac{dN}{d\log_{10}L} \\
 &=\phi_{\star}\left(\frac{L}{L_{\star}}\right)^{1-\beta}
 \exp \left[-\frac{1}{2 \gamma}^{2} \log ^{2}\left(1+\frac{L}{L_{\star}}\right)\right],
\end{aligned}
\end{eqnarray}
where $L_{\star}$ determines the location of the  knee in the LF, $\beta$ and $\gamma$ fit the faint and bright ends of the LF, respectively, and $\Phi_{\star}$ is used for the normalization.
In this work, we consider three LF models, all of which adopt the same LE function:
\begin{eqnarray}
        \label{e2A}
        e_2(z)=(1+z)^{k_{1} + k_{2} z}.
\end{eqnarray}
The DE function $e_1(z)$ has three different forms depending on the model: $e_1(z)=1$ for model A, an exponential form
\begin{eqnarray}
        \label{e1B}
        e_1(z)=10^{p_1 z},
\end{eqnarray}
for model B, and
\begin{eqnarray}
        \label{e1C}
        e_1(z)=(1+z)^{p_{1} + p_{2} z},
\end{eqnarray}
for model C. In the above equations, $k_{1}$, $k_{2}$, $p_{1}$, and $p_{2}$ are free parameters. Model A is the pure luminosity evolution (PLE) model, which is the most commonly used model for the radio LFs of SFGs in the literature \citep[e.g.,][]{2017A&A...602A...5N}. Models B and C can be referred to as the mixture evolution \citep[e.g.,][]{yuan2016mixture,yuan2017mixture} or luminosity and density evolution \citep[LADE, e.g.,][]{2010MNRAS.401.2531A} models.

\subsection{Model selection}

In order to evaluate which model is a better fit to the data, a helpful tool is the information criterion\citep{Takeuchi:1999cu}. The Akaike Information Criterion \citep[AIC;][]{1974ITAC...19..716A} is one of the most widely used information criterion. For our problem, the AIC can be written as:
\begin{eqnarray}
        \label{aic}
        \text{AIC}= S_{\mathrm{all}}(\hat{\theta}) + 2q,
\end{eqnarray}
where $S_{\mathrm{all}}$ is given in Equation (\ref{chi2all}), $\hat{\theta}$ is the best-fit model parameters, and $q$ is the number of parameters for each model. The model with the smallest value of AIC is considered to be the most accurate. Another commonly used criterion is the Bayesian information criterion \citep[BIC,][]{Schwarz1978}, which can be written as
\begin{eqnarray}
        \label{bic}
        \text{BIC}(q)= S_{\mathrm{all}}(\hat{\theta}) + q\text{ln} ~ n,
\end{eqnarray}
where $n$ is the sample size. When calculating the $S_{\mathrm{all}}$ values for our three models, the weight factor $A_0$ in Equation (\ref{chi2all}) is set to $1$. The AIC and BIC values are listed in Table \ref{aicbicpara}. We find that the AIC and BIC are consistent with each other, both indicating that the LADE model is superior to the PLE model. The AIC value of Model B is slightly smaller than that of Model C, implying that Model B could be taken as our preferred model.

\begin{figure*}
        \centering
        \includegraphics[width=\textwidth]{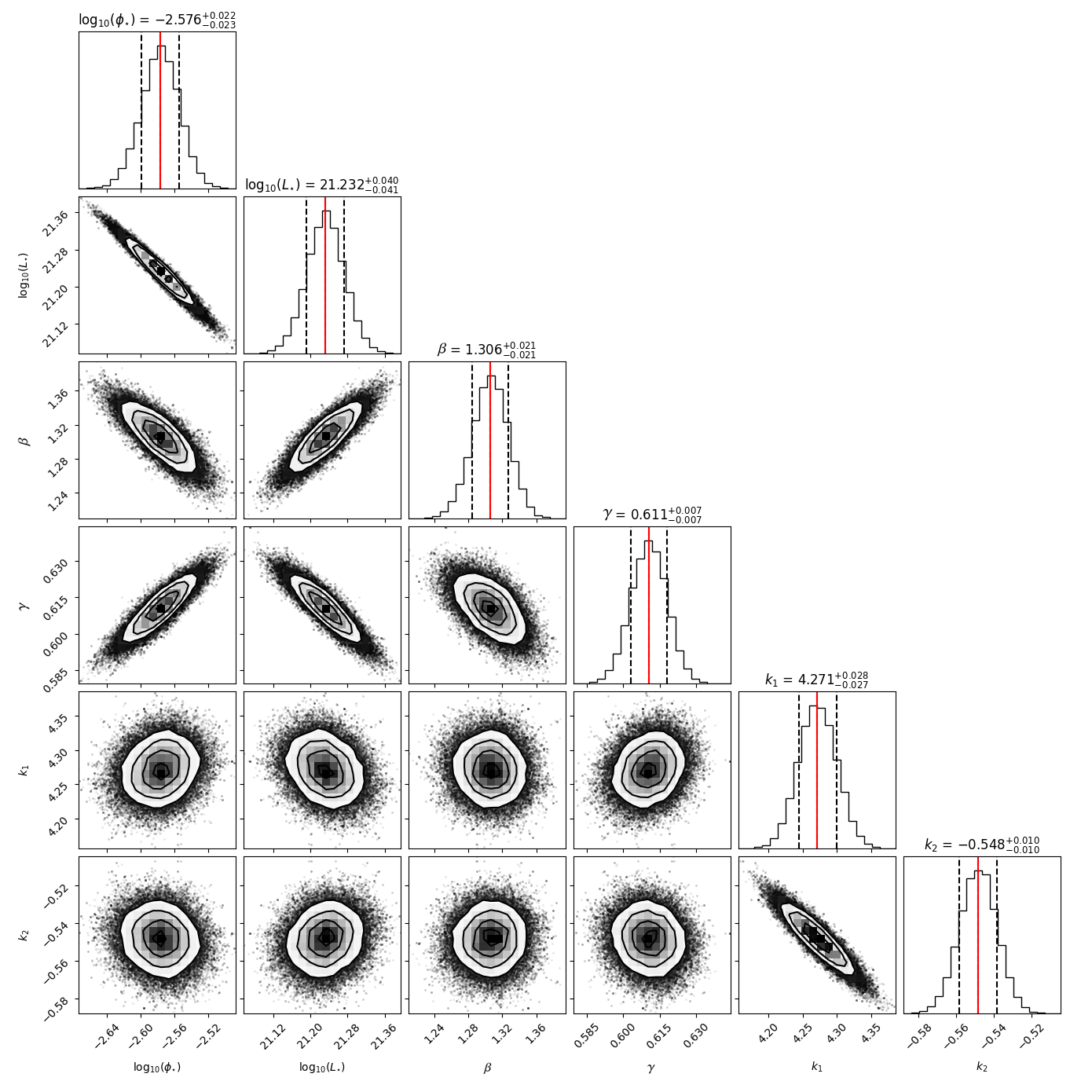}
        \caption{
                Corner plot showing the one- and two-dimensional projections of the posterior probability distributions of the parameters for Model A obtained from the MCMC run. The histograms on the diagonal show the marginalized posterior densities for each parameter (vertical dashed lines denote the 16th and 84th percentiles). The off-diagonal panels show the 2D joint posterior densities of all couples of parameters, with 1$\sigma$, 2$\sigma$, and 3$\sigma$ contours shown by black solid lines. Our best-fitting parameters are marked by red vertical solid lines.}
        \label{fig:cornerplotA}
\end{figure*}

\begin{figure*}
        \centering
        \includegraphics[width=\textwidth]{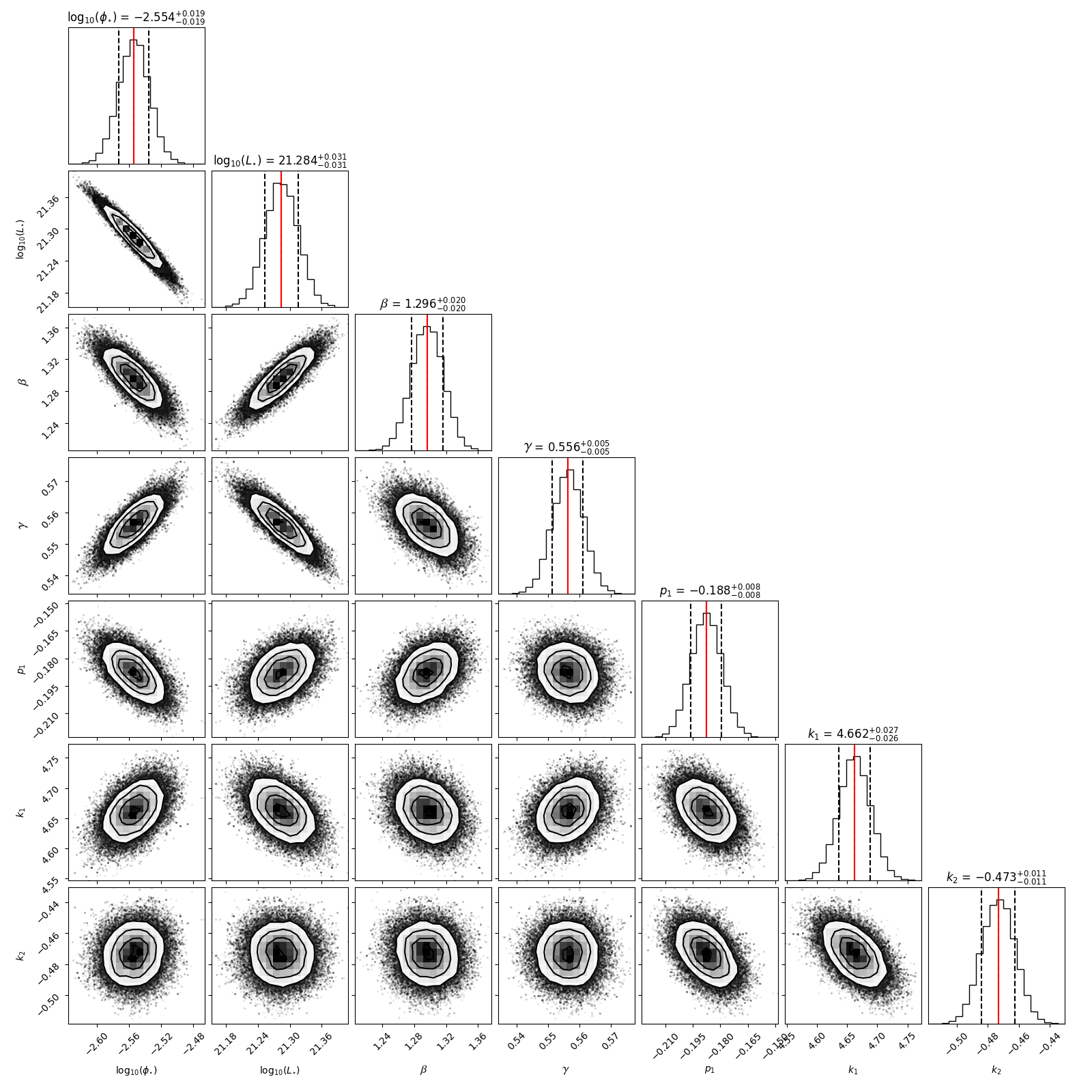}
        \caption{Similar to Figure \ref{fig:cornerplotA}, but for Model B.}
        \label{fig:cornerplotB}
\end{figure*}

\begin{figure*}
        \centering
        \includegraphics[width=\textwidth]{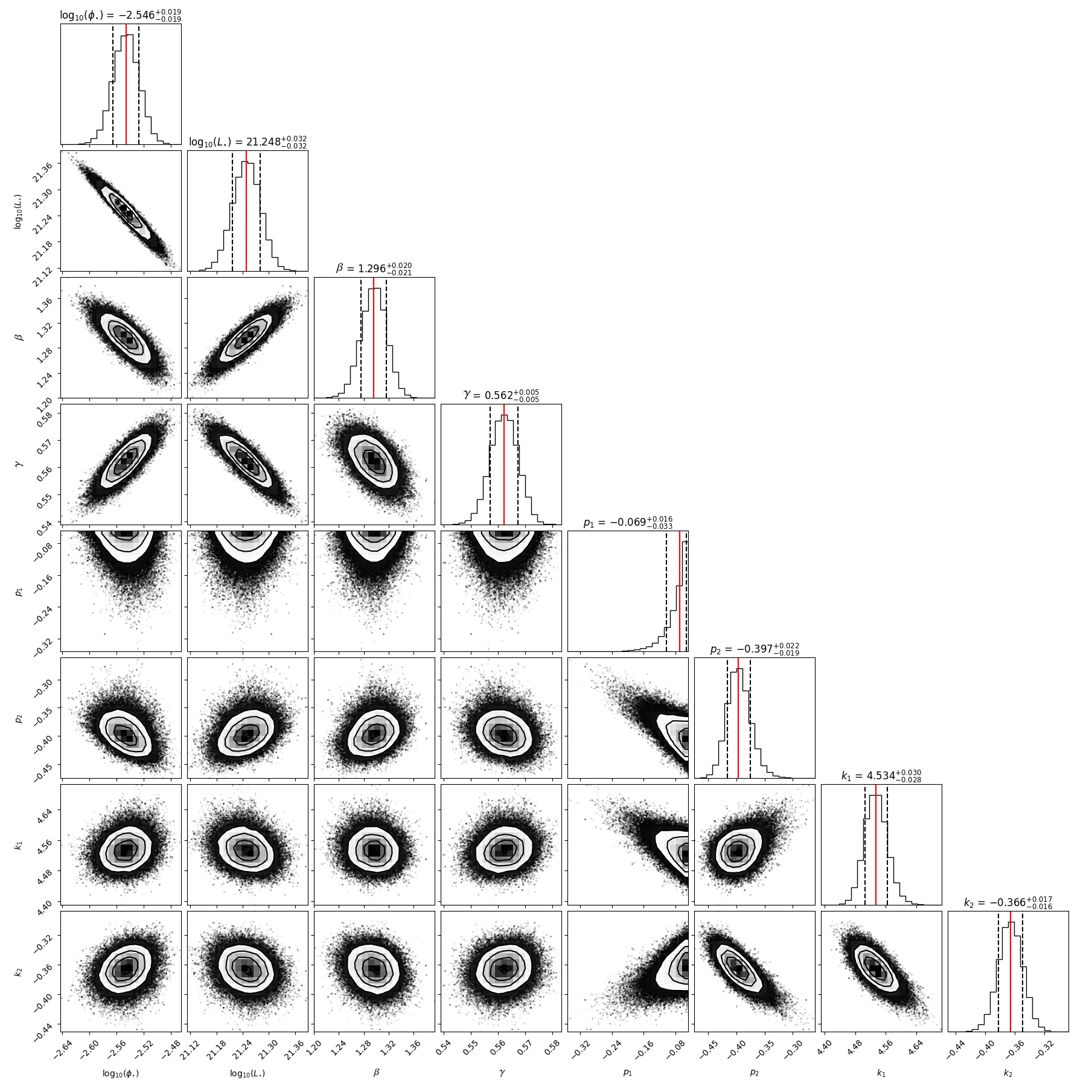}
        \caption{Similar to Figure \ref{fig:cornerplotA}, but for Model C.}
        \label{fig:cornerplotC}
\end{figure*}

\section{Results}
\label{results}

\begin{figure*}
\centering
\includegraphics[width=\textwidth]{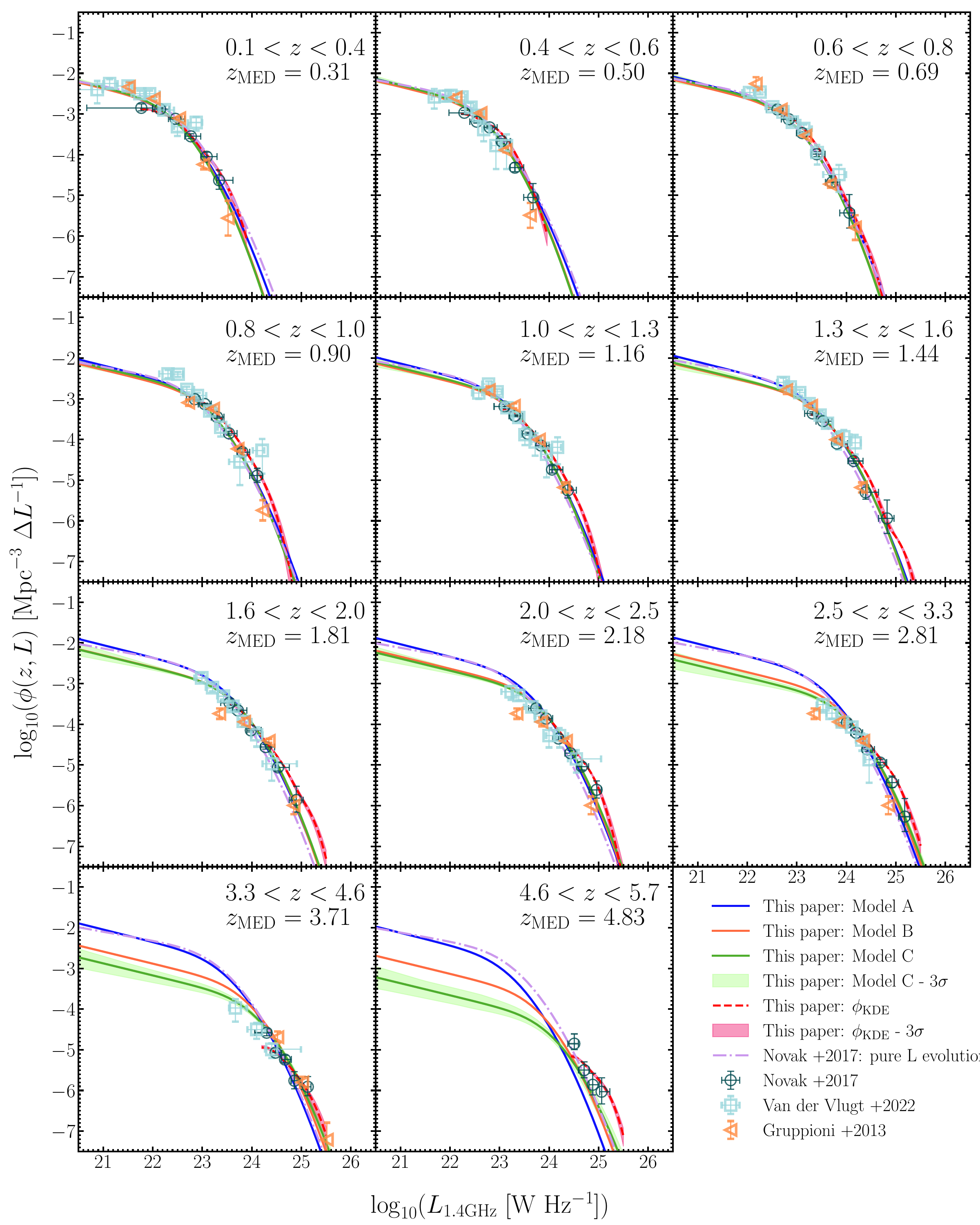}
\caption{
Radio LFs of SFGs at various redshifts compared with the previous estimates specified in the inset. The best-fit LFs for Models A, B, and C in each redshift bin are shown by the blue, orange, and green solid lines, respectively. The light green shaded area shows the 3$\sigma$ confidence interval for Model C.
The red dashed curves represent the KDE LFs (see Appendix \ref{KDEform}), with the 3$\sigma$ confidence interval shown by the pink shaded area. The purple dash-dotted line depicts the PLE model of \cite{2017A&A...602A...5N}. We also compare with the binned LFs from \cite{Gruppioni_2013}, \cite{2017A&A...602A...5N}, and \cite{2022ApJ...941...10V}.}

\label{fig:PLFTimeRoman}
\end{figure*}

\begin{table*}
\setlength{\abovecaptionskip}{0cm} %
\setlength{\belowcaptionskip}{0cm}
\renewcommand{\arraystretch}{1.7} %
\caption{Best-fit parameters for models A, B, and C}
\begin{center}
\resizebox{\textwidth}{15mm}{ %
\begin{tabular}[width=\columnwidth]{lcccccccc}
\hline\hline
Model~~ & $\text{log}_{10}(\phi_{\star})$ & $\text{log}_{10}(L_{\star})$
& $\beta$ & $\gamma$ & $k_1$ & $k_2$ & $p_1$ & $p_2$\\
\hline
A~~ & $-2.576_{-0.023}^{+0.022}$ & $21.232_{-0.041}^{+0.040}$ & $1.306_{-0.021}^{+0.021}$ & $0.611_{-0.007}^{+0.007}$ & $4.271_{-0.027}^{+0.028} $ & $-0.548_{-0.010}^{+0.010}$ & $\ldots$ & $\ldots$\\
B~~ & $-2.554_{-0.019}^{+0.019}$ & $21.284_{-0.031}^{+0.031}$ & $1.296_{-0.020}^{+0.020}$ & $0.556_{-0.005}^{+0.005}$ & $4.662_{-0.026}^{+0.027} $ & $-0.473_{-0.011}^{+0.011}$ & $-0.188_{-0.008}^{+0.008}$ & $\ldots$\\
C~~ & $-2.546_{-0.019}^{+0.019}$ & $21.248_{-0.032}^{+0.032}$ & $1.296_{-0.021}^{+0.020}$ & $0.562_{-0.005}^{+0.005}$ & $4.534_{-0.028}^{+0.030} $ & $-0.366_{-0.016}^{+0.017}$ & $-0.069_{-0.033}^{+0.016}$ & $-0.397_{-0.019}^{+0.022}$\\
\hline
\end{tabular}}
\end{center}
\text{Note. Units --- $\phi_{\star}$: [${\rm Mpc^{-3} dex^{-1}}$],\,\, $L_{\star}$: [${\rm W Hz^{-1}}$]. The best-fit parameters as well as their 1$\sigma$ errors for models A, B and C.}
\label{modelpara}
\end{table*}

\begin{table}
\setlength{\abovecaptionskip}{0cm} 
\setlength{\belowcaptionskip}{0cm}
\renewcommand{\arraystretch}{1.1} %
\caption{Values of AIC and BIC for models A, B, and C}
\begin{center}
\resizebox{0.7\columnwidth}{!}{ 
\small
\begin{tabular}{lcc} 
\hline\hline
Model~~ & AIC & BIC \\
\hline
 A~~ & 115258.9 & 115298.9 \\
 B~~ & 114459.7 & 114506.5 \\
 C~~ & 114560.5 & 114613.9 \\
\hline
\end{tabular}}
\end{center}
\label{aicbicpara}
\end{table}

\subsection{Analytical LFs}
\label{RLFs}
The parameters in our model LFs are estimated via the MCMC algorithm, which is
performed using the Python package {\sc{emcee}} of \citet{foreman2013emcee}.
The {\sc{emcee}} algorithm improves the exploration of the parameter space by using an ensemble of chains with different initial conditions. This approach helps to avoid local minima and ensures a more comprehensive exploration of the parameter space. We assume uniform priors on all the parameters. The marginalized one- and two-dimensional posterior probability distributions  of the parameters for Models A, B, and C are shown in Figures \ref{fig:cornerplotA}, \ref{fig:cornerplotB}, and \ref{fig:cornerplotC}, respectively. These corner plots illustrate that all the parameters for our three models are well constrained. Table \ref{modelpara} reports the best-fit parameters and their 1$\sigma$ uncertainties for the three models.

Figure \ref{fig:PLFTimeRoman} shows our best-fit LFs for Models A (solid blue lines) , B (solid orange lines), and C (solid green lines).
All the LFs are measured at the rest-frame 1.4 $\rm GHz$.
We also compare our result with the binned LFs from \cite{Gruppioni_2013}, \cite{2017A&A...602A...5N}, and \cite{2022ApJ...941...10V}, which are represented by orange left-pointing triangles, dark blue circles, and sky blue squares with error bars, respectively. At lower redshifts of $z<1.0$, our three models are barely distinguishable. As redshift increases, Model A begins to diverge, predicting larger number densities at $L <L_{\star}$ than models B and C. This deviation increases towards higher redshifts, and the disagreement is in excess of the 3$\sigma$ confidence intervals at $z>\sim1$. We also note that, at $z>\sim2,$ the binned LFs of \cite{2022ApJ...941...10V} present a decline in number density at the faint end, while model A cannot reproduce the behavior. This indicates that model A is not applicable to describe the evolution of LFs at higher redshift. By contrast, Models B and C are in good agreement with the binned LFs for all redshift intervals.

In Figure \ref{fig:PLFTimeRoman}, the purple dotted lines depict the PLE model of \cite{2017A&A...602A...5N}, which is generally in agreement with our Model A. Nevertheless, the difference between the two results increases with redshift. This is not surprising given that the PLE model of \cite{2017A&A...602A...5N} is constrained through simultaneously fitting the LF points in all redshift bins, while these LF points are estimated using the traditional $1/V_{\rm max}$ method. Although their modeling result would be dependent on the estimation accuracy of $1/V_{\rm max}$. Our analytical LFs are obtained through a full maximum-likelihood analysis, and are therefore independent of the $1/V_{\rm max}$ estimates.

\begin{figure}
        \hspace{-0.2in}
        \includegraphics[width=0.55\textwidth]{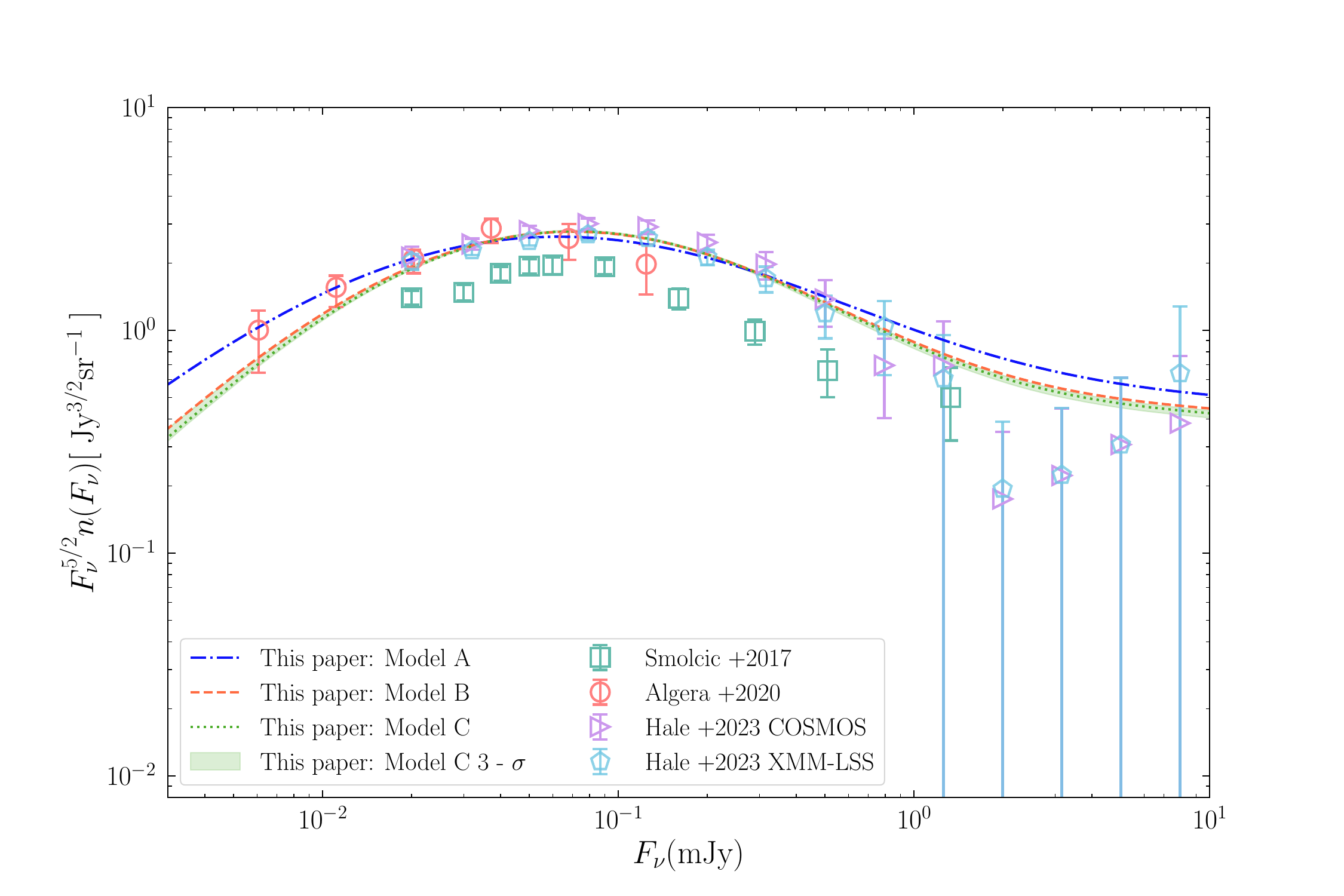}
        \caption{Comparison of our best-fit models with the Euclidean normalized $1.4$ GHz source counts for SFGs observed in the literature. The blue dash-dotted line, orange dashed line, and green dotted line show our best-fit source counts of Models A, B, and C, respectively. The source counts from \cite{2023MNRAS.520.2668H} in the COSMOS and XMM-LSS fields are shown as purple right triangles and light blue pentagons, respectively. Also shown are the observed source counts from \cite{2017A&A...602A...1S} (green squares), \cite{2020ApJ...903..139A} (orange circles).}
        \label{fig:Source Counts}
\end{figure}

\begin{figure*}
        \centering
        \includegraphics[width=0.5\textwidth]{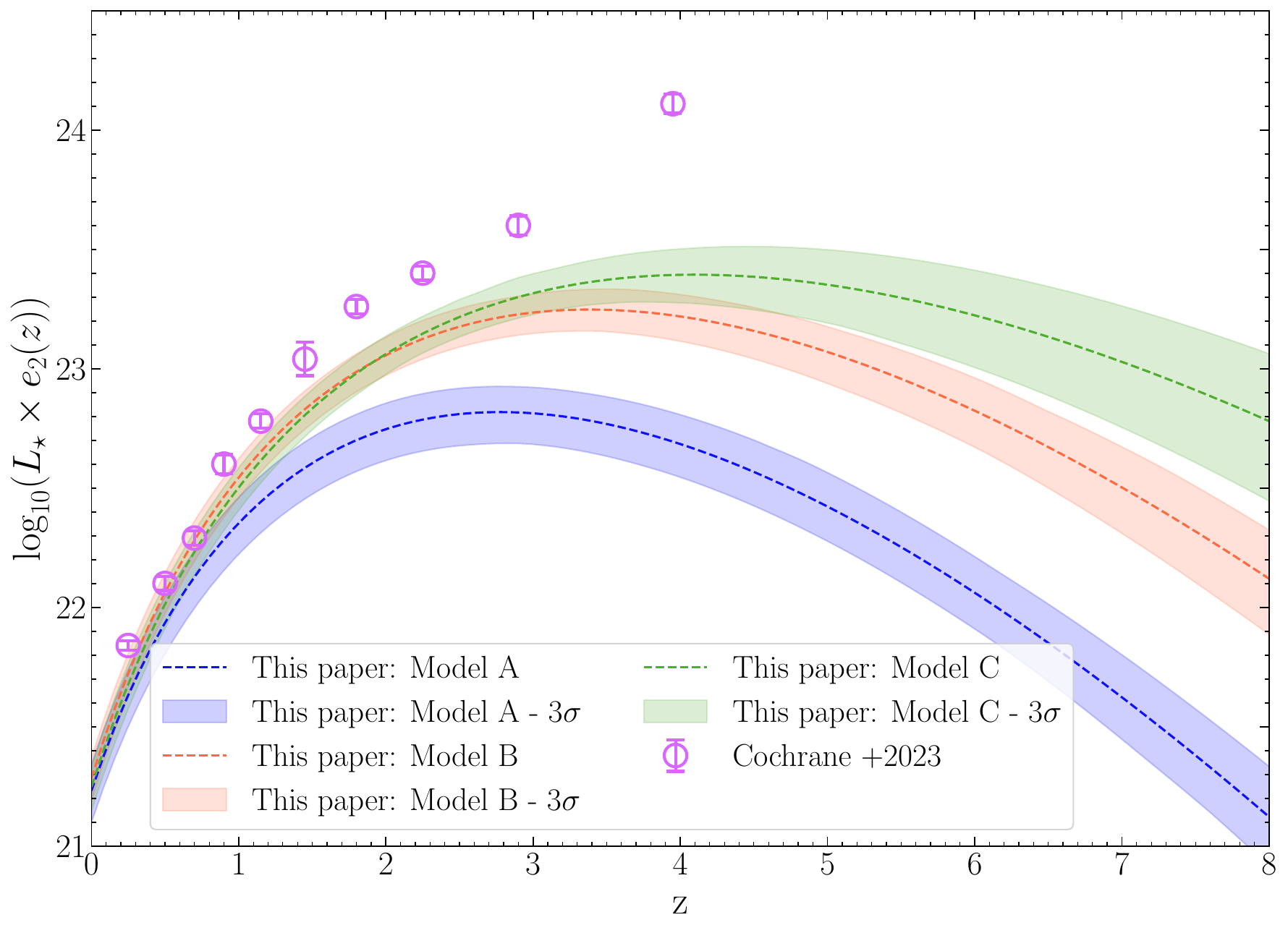}
        \hspace{-0.1in}
        \includegraphics[width=0.5\textwidth]{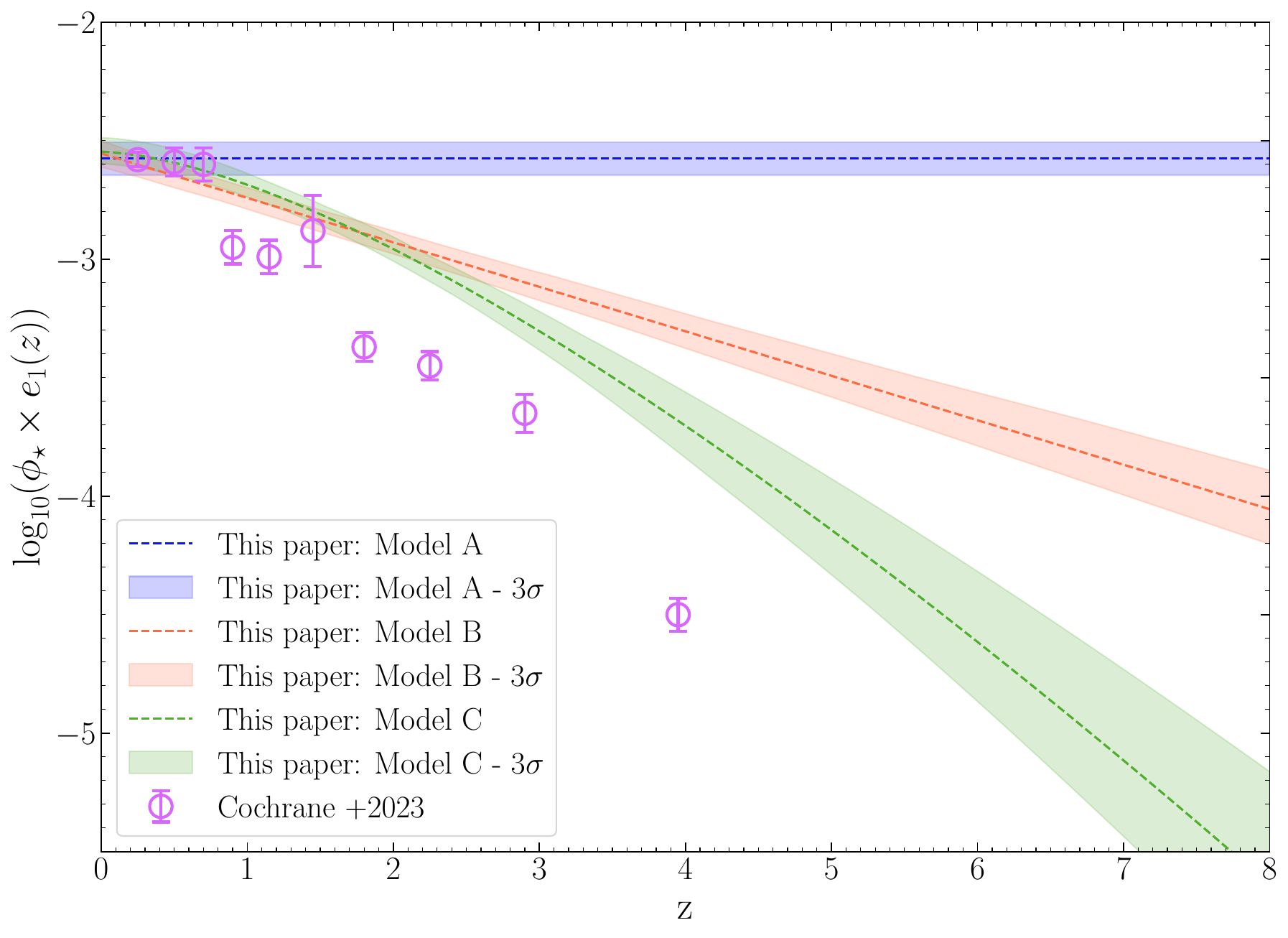}
        \caption{$L_{\star}\times e_2(z)$ (left panel) and $\phi_{\star}\times e_1(z)$ (right panel) for our three LF models shown in different colours, compared with the inference (in all fields) from \citet{2023MNRAS.523.6082C}. The light shaded areas take into account the 3$\sigma$ error bands. Our $L_{\star}$ and $\phi_{\star}$ have been converted to $150$ MHz by assuming a spectral index of 0.7. The purple circles represent the variation of $L_{\star}$ (left panel) and $\phi_{\star}$ (right panel) as functions of redshift inferred by \citet{2023MNRAS.523.6082C}.}
        \label{fig:evolution}
\end{figure*}

\subsection{Fitting the observed source counts}
We calculate the source counts for each of our three model LFs using Equation(\ref{sc}). The result is shown in Figure \ref{fig:Source Counts} as blue dash-dotted line, orange dashed line, and green dotted line for models A, B, and C, respectively. In the Figure, we compare our models with the Euclidean normalized $1.4$ GHz source counts for SFGs measured from \cite{2023MNRAS.520.2668H}, \cite{2017A&A...602A...1S}, and \cite{2020ApJ...903..139A}. All our three models can reproduce the \cite{2020ApJ...903..139A} and \cite{2023MNRAS.520.2668H} source counts fairly well, but the \cite{2017A&A...602A...1S} result is systematically lower than the others. The discrepancies could be caused by multiple factors, such as field-to-field variation, differences in the assumptions used to calculate completeness, and resolution bias \cite[see][]{2023MNRAS.520.2668H}.
We note that, at the bright region ($F_{\nu} > 1$ mJy), the measurements of \cite{2023MNRAS.520.2668H} display huge uncertainties. This is due to the contamination of AGNs at these fluxes, preventing an easy classification between SFGs and AGNs.

\subsection{LADE versus PLE}
Due to the limitation of survey depth, most of the existing radio studies barely reach the knee of the SFG LF at z > 1. If fitting the LF using a LADE model, the DE and LE parameters may become degenerate \citep[also see][]{2022ApJ...941...10V}. On several occasions, this has led previous authors to assume a PLE model for their SFG LFs. By combining data from
the ultradeep COSMOS-XS survey and the shallower VLA-COSMOS 3 GHz large project, \citet{2022ApJ...941...10V} was able to jointly constrain the LE and DE, finding evidence for significant DE. From Figure \ref{fig:PLFTimeRoman}, we also find that the LADE model is superior to the PLE model.

Very recently, \citet{2023MNRAS.523.6082C} measured the 150 MHz LFs of SFGs using data from the Low Frequency Array (LOFAR) Two Metre Sky Survey in three well-studied extragalactic fields, Elais-N1, Bo$\ddot{o}$tes, and the Lockman Hole.
By fixing the faint and bright end shape of the radio LF to the local values (equivalent to fixing $\gamma=0.49$ and $\beta=1.12$ in Equation \ref{rho}), these latter authors fitted the LF points (via $1/V_{\rm max}$) to find the best-fit $L_{\star}$ and $\phi_{\star}$ at each individual redshift bin. Their Figure 7 shows the variation of $L_{\star}$ and $\phi_{\star}$ as functions of redshift. The method of \citet{2023MNRAS.523.6082C} is equivalent to assuming a LADE model for their LFs. The variation of $L_{\star}$ and $\phi_{\star}$ as functions of redshift in their analysis correspond to $L_{\star}\times e_2(z)$ and $\phi_{\star}\times e_1(z)$ in our work. In Figure \ref{fig:evolution}, we show $L_{\star}\times e_2(z)$ and $\phi_{\star}\times e_1(z)$ for our three LF models compared with the inference from \citet{2023MNRAS.523.6082C}.
We converted our $L_{\star}$ to $150$ MHz by assuming a spectral index of 0.7.
Below $z<1.5$, our two LADE models ---especially Model C--- agree well with the $L_{\star}$ and $\phi_{\star}$ evolutions given by \citet{2023MNRAS.523.6082C}.
Above $z>1.5$, the $L_{\star}$ evolution curve obtain by these authors is significantly higher than those of our models (outside the 3$\sigma$ uncertainties), while their $\phi_{\star}$ evolution falls more rapidly than those of our models out to high redshift.

The discrepancies could be explained as follows: At any redshift bin, the best-fit $L_{\star}$ and $\phi_{\star}$ are negatively correlated with each other (e.g., see the right panel of Figure 4 in \citet{2023MNRAS.523.6082C}), implying that the LE and DE parameters are degenerate. The degeneracy should be stronger at higher redshift where the knee location of LFs is increasingly difficult to identify. We highlight the fact that \citet{2023MNRAS.523.6082C} fitted the LF points (via $1/V_{\rm max}$) for each redshift bin individually. At higher redshift, the LF points estimated by $1/V_{\rm max}$ usually have larger uncertainty. Fitting these discrete LF points to find a precise knee location would be very difficult. This will inevitably bias the inferred $L_{\star}$ and $\phi_{\star}$ values. Therefore, their $L_{\star}$ and $\phi_{\star}$ evolutions for high redshift are subject to uncertainties due to dual factors. Our LADE modeling is also subject to the uncertainty due to degeneracy, but is free from the $1/V_{\rm max}$ estimates.

\citet{2023MNRAS.523.6082C} found that their $\phi_{\star}$ remains roughly constant back to $z\sim0.8$ but then falls steeply at higher redshifts.
Our Model C displays a similar trend. A comparison between the inference of \citet{2023MNRAS.523.6082C}  and our models lends strong support to the efficacy of our LADE models.

\begin{figure*}
        \centering
        \includegraphics[width=0.49\textwidth]{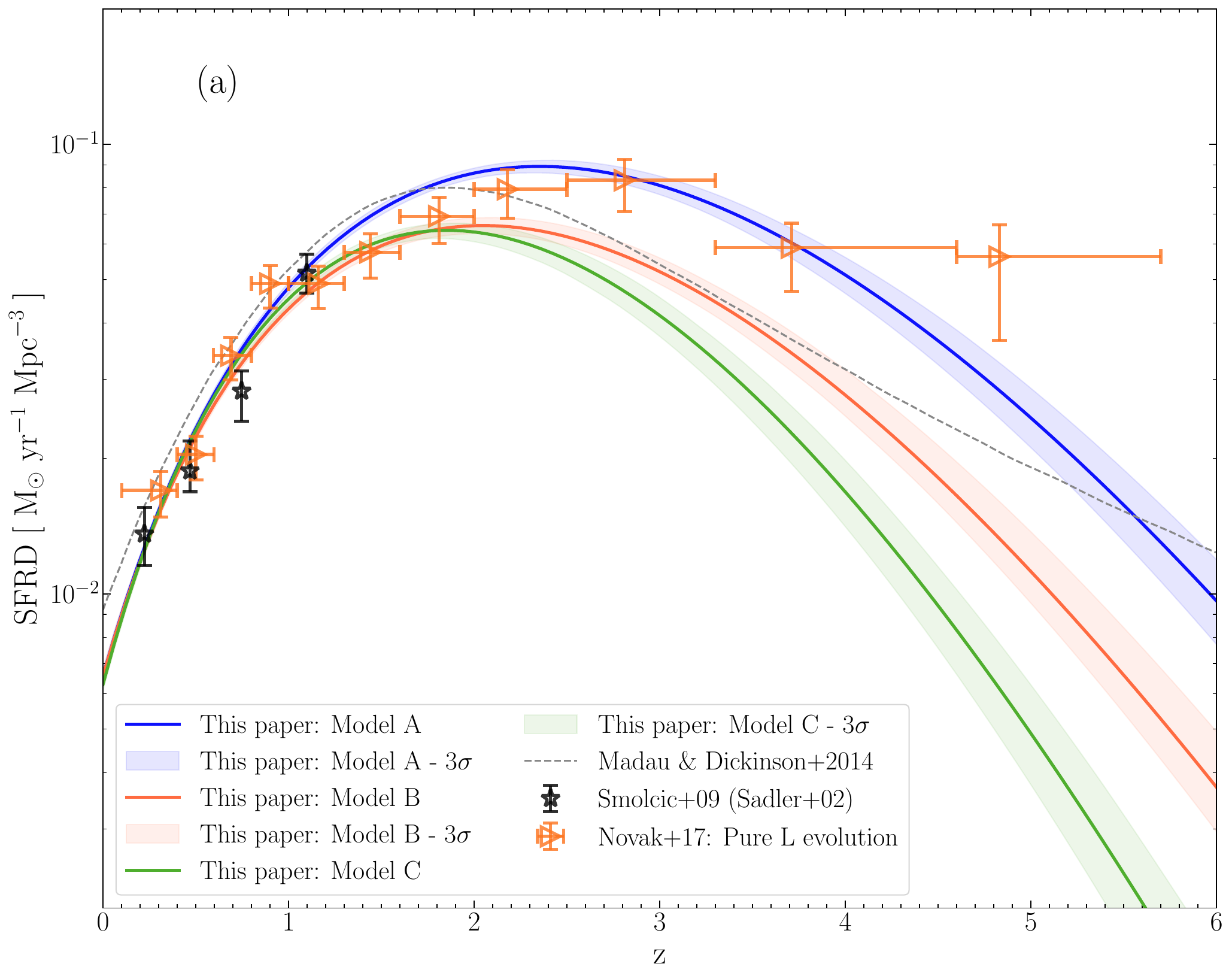}
        \includegraphics[width=0.49\textwidth]{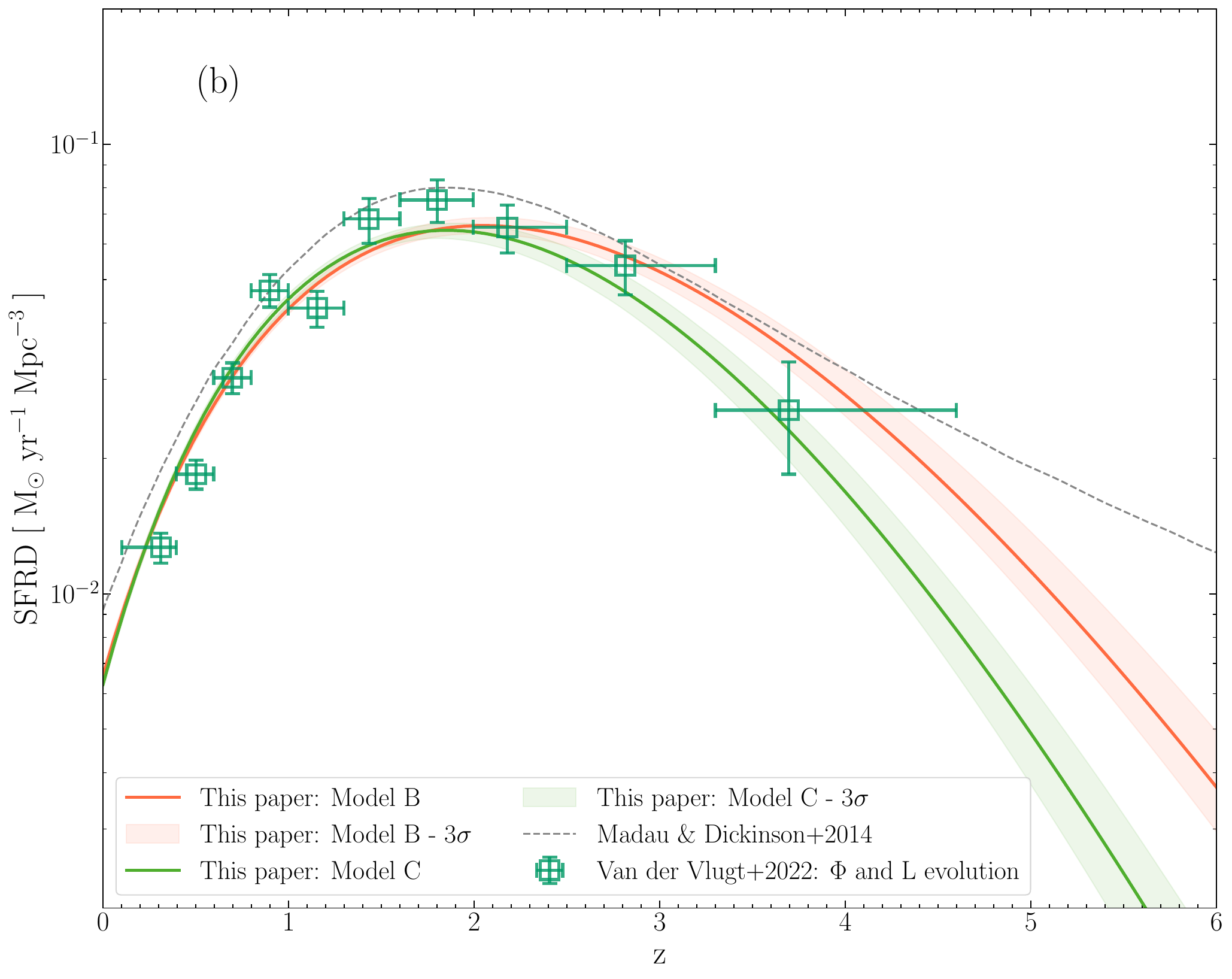}
        \includegraphics[width=0.49\textwidth]{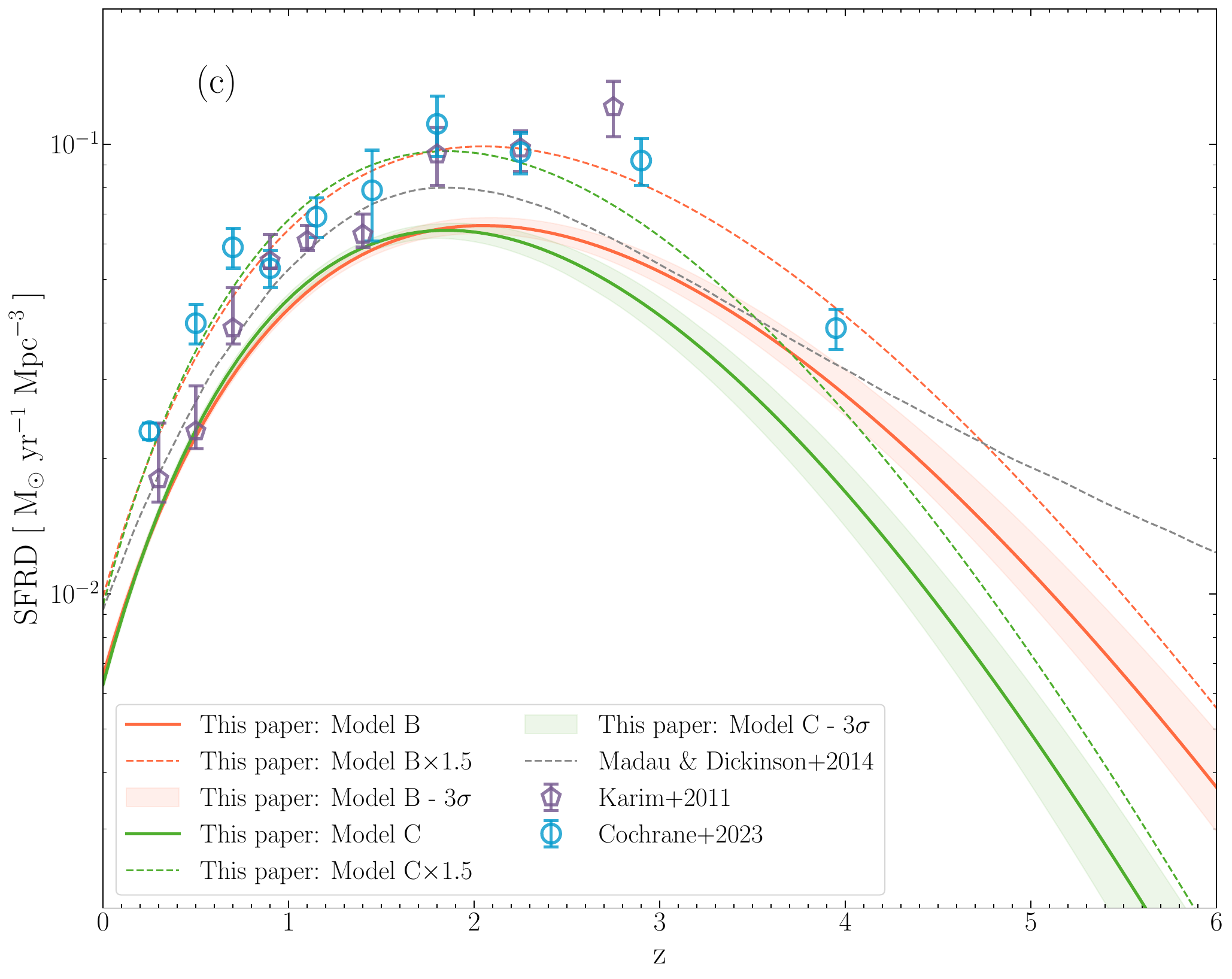}
    \includegraphics[width=0.49\textwidth]{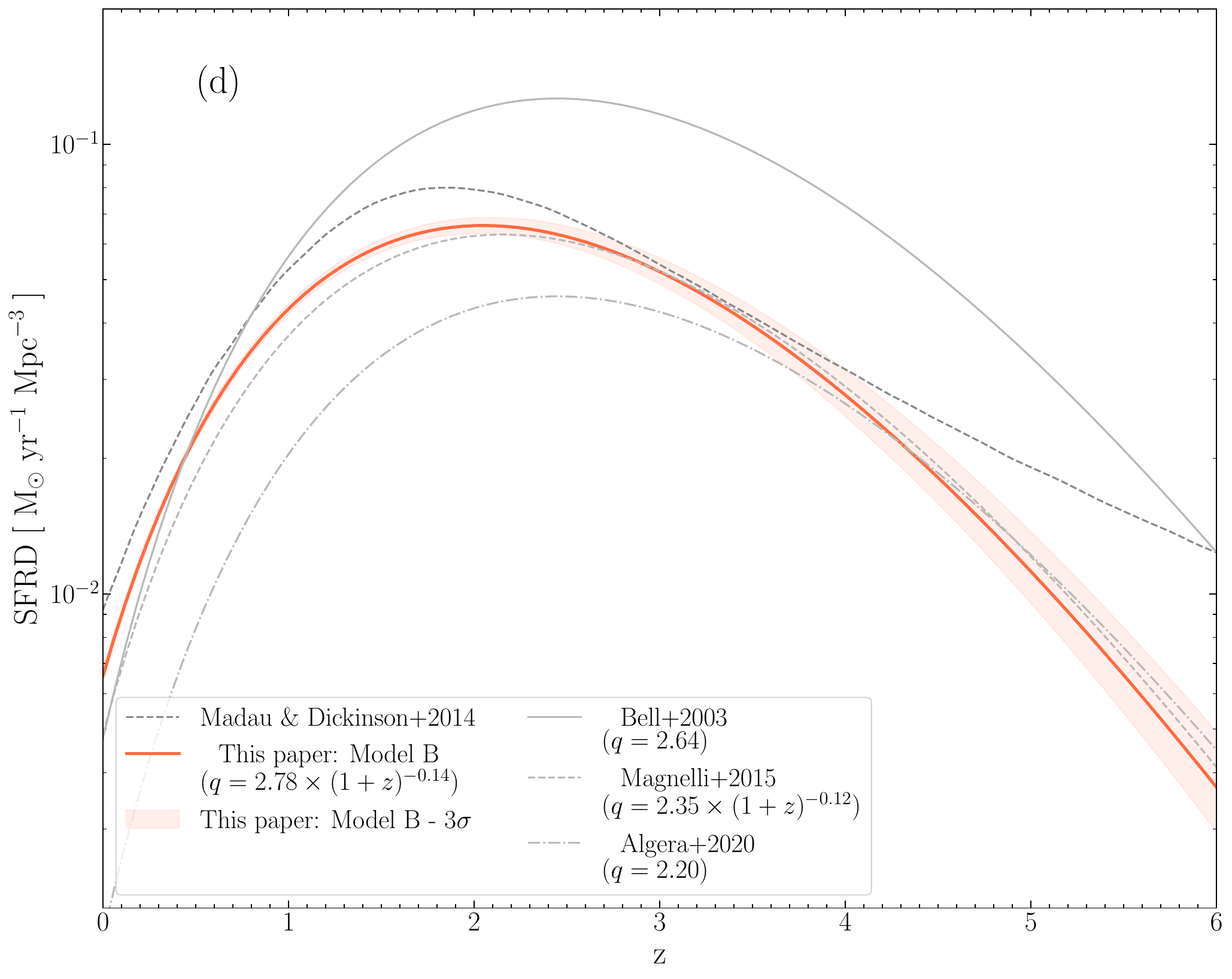}
        \caption{History of the cosmic SFRD. Our SFRD history results are shown as blue, orange, and green solid lines for Models A, B, and C, respectively. The light shaded areas take into account the $3\sigma$ error bands. The compilation of \citet{Madau_2014} is shown as a black dashed line in all panels. All data shown for comparison are indicated in the legend of each panel; see text for details.}
        \label{fig:SFRD}
\end{figure*}

\section{ History of the cosmic star formation rate density}
\label{Cosmic star formation rate density history}

\subsection{Calculating the SFRD}
Now that we have obtained the rest-frame $\mathrm{1.4~GHz}$ LF, we can investigate how the SFRD evolves with redshift. We use the functional form provided in \citet{2021A&A...647A.123D} to convert the radio luminosity into an SFR:
\begin{align}
\frac{\mathrm{SFR}}{M_{\odot}\mathrm{yr}^{-1}}=f_{\mathrm{IMF}}\times10^{-24}10^{q_{\mathrm{IR}}(z)}\frac{L_{1.4\mathrm{GHz}}}{\mathrm{WHz}^{-1}},
        \label{eq:SFR}
\end{align}
where $f_{\text{IMF}}$ is a factor accounting for the IMF  ($f_{\text{IMF}}=1$ for a \citet{Chabrier_2003} IMF and $f_{\text{IMF}}=1.7$ for a \citet{1955ApJ...121..161S} IMF), and $L_{1.4\mathrm{GHz}}$ is rest-frame $1.4\,\mathrm{GHz}$ radio luminosity. Following \citet{2017A&A...602A...5N}, we use the Chabrier IMF in the following analysis.
In Equation (\ref{eq:SFR}), $q_{\mathrm{IR}}(z)$ is the FIR-to-radio luminosity ratio, which is conventionally used to parametrize the FIR--radio correlation in SFGs, and is defined as
\begin{align*}
q_{\mathrm{IR}}=\log\left(\frac{L_{\mathrm{IR}}\left[\mathrm{W}\right]}{3.75\times10^{12}\left[\mathrm{Hz}\right]}\right)-\log(L_{1.4\mathrm{GHz}}[\mathrm{W\,Hz^{-1}}]),
\end{align*}
where $L_{\mathrm{IR}}$ is the total IR luminosity (rest-frame 8-1000 $\mu \text{m}$), and $3.75\times10^{12} \mathrm{Hz}$ represents the central frequency over the far-infrared domain.

Although $q_{\mathrm{IR}}$ is typically taken to be a constant value derived for local galaxies, recent observations suggest that the $q_{\mathrm{IR}}$ value probably changes with redshift \citep[e.g.,][]{Sargent_2010, magnelli2015far,Delhaize_2017,Calistro_2017}. Recently, for the first
time, \citet{2021A&A...647A.123D} calibrated $q_{\mathrm{IR}}$ as a function of both stellar mass ($M_{\star}$) and redshift. These latter authors found that $q_{\mathrm{IR}}$ primarily evolves with $M_{\star}$, and only weakly with redshift.
This finding implies that using radio emission as an SFR tracer requires $M_{\star}$-dependent conversion factors, but its robustness still needs to be verified with further study. In this work, we use the expression given by \citet{2017A&A...602A...5N}:
\begin{align}
        q_{\mathrm{IR}}(z)=(2.78 \pm 0.02) \times(1+z)^{-0.14 \pm 0.01},
        \label{eq:qtir}
\end{align}
which is the updated version of the analysis of \citet{Delhaize_2017}, using a new sample selection criteria to exclude AGN.

The SFRD of a given epoch can then be estimated by the following integral:
\begin{align}
        \text{SFRD} = \int_{L_{\text{min}}}^{L_{\text{max}}} \Phi(L,z) \times \text{SFR}(L_{1.4\,\text{GHz}}) \, \mathrm{d} \log_{10} L \,.
        \label{eq:SFRD}
\end{align}
To obtain the SFRD for a given epoch, we performed a numerical integration of the analytical form of the LF in each redshift bin, employing the best-fit evolution parameters presented in Table \ref{modelpara} and Figure \ref{fig:PLFTimeRoman}. The resulting integral provides an estimate of the SFRD. Unless otherwise specified, our reported SFRD values correspond to the integral of the fitted LF from 0.0 to $\to \infty$. In Figure \ref{fig:SFRD}(a), our SFRD history results are shown as blue, orange, and green solid lines for Models A, B, and C, respectively. The light shaded areas take into account the $3\sigma$ error bands. Our three models coincide at lower redshift of $z<1$. Model A gradually separates from the other two models towards higher redshift. Models B and C start to diverge at $ z > \sim 2$, but the difference is within the $3\sigma$ confidence intervals.

\subsection{Comparison with the literature}

In Figure \ref{fig:SFRD}, we show the SFRD history derived from our three models compared with those in the literature derived at different wavelengths.
The SFRD from the review by \citet{Madau_2014}, who performed a fit on a collection of previously published UV and IR data, is shown as the black dashed curve in all panels for ease of comparison. From Figure \ref{fig:SFRD}(a),
we find that below $z \sim 1.5$, the rate at which the SFRD increases with redshift in our three models shows striking similarity to the trend observed by \citet{Madau_2014}, although their SFRD is slightly higher.
The discrepancy in magnitude is probably due to the assumed FIR–radio relation in our result, which calibrates radio emission as a tracer of SFR.
Above $z \sim 1.5$, our PLE model predicts a significantly higher SFRD, while the SFRD given by our LADE models is lower than that of \citet{Madau_2014}. The SFRD derived from all our three models turns over at a slightly higher redshift ($2<z<2.5$) and falls more rapidly than that of \citet{Madau_2014} out to high redshift. A similar behavior was observed in recent radio estimates by \citet{2022ApJ...941...10V}.

In Figure \ref{fig:SFRD}(a), we also show our Model A SFRD compared to the radio estimates from \citet{2009ApJ...690..610S} and \citet{2017A&A...602A...5N}. \citet{2009ApJ...690..610S} derived the SFRD out to $z = 1.3$ by assuming a PLE LF and a nonevolving FIR–radio correlation established by \citet{Bell_2003}.
We find an agreement with the \citet{2009ApJ...690..610S} estimates within the permissible error ranges, and therefore their result provides a good consistency check for our models at low redshift. The result of \citet{2017A&A...602A...5N}, who also assumed a PLE LF, is the key comparative object for our model A, because our analysis is based on the sample studied by these latter authors. Overall, the curve of our Model A SFRD seems to be a good fit to their SFRD points. Nevertheless, there are two points at $z<2$ and one point at $z>2$ that seem to disagree with our Model A at the 3$\sigma$ level. Because  we use the same $q_{\mathrm{IR}}(z)$ evolution to calculate the SFRD as \citet{2017A&A...602A...5N} did, any discrepancy between the two results can only arise from the difference in LF (see Figure \ref{fig:PLFTimeRoman}). As discussed in section \ref{RLFs}, the analytical LFs of \cite{2017A&A...602A...5N} are obtained by fitting the $1/V_{\rm max}$ LF points in all redshift bins, while our LFs are obtained through a full maximum-likelihood analysis, incorporating additional constraints from source counts and local LFs. Therefore, our LFs should be more accurate than that of \cite{2017A&A...602A...5N}, making our SFRD an improvement on their estimates.

In Figure \ref{fig:SFRD}(b), we show the SFRD derived from our Models B and C compared to the radio estimates from  \citet{2022ApJ...941...10V}. These latter authors also assumed a LADE LF ---similar to our Models B and C--- to calculate their SFRD based on the combined data set from the ultradeep COSMOS-XS survey and the VLA-COSMOS 3 GHz large project. Although adopting a different $q_{\mathrm{IR}}(z)$ evolution from that used here, their result is in good agreement with those of our Models B and C ---especially our Model C--- within the error bars. We note that three points of the estimates of \citet{2022ApJ...941...10V} show a slightly elevated SFRD compared to our model prediction. The discrepancies could be attributed to the different $q_{\mathrm{IR}}(z)$ evolution used by these authors, or the uncertainties in their LF measurement
propagated from the $1/V_{\rm max}$ estimator. Similar to \citet{2017A&A...602A...5N}, \citet{2022ApJ...941...10V} also obtained their analytical LFs by fitting the $1/V_{\rm max}$ LF points in all redshift bins.

Figure \ref{fig:SFRD}(c) shows the SFRD derived from our Model B compared to the radio estimates from \citet{karim2011star} and \citet{2023MNRAS.523.6082C}. Our estimates are slightly lower than that of \citet{karim2011star}, with the difference increasing with redshift.
The discrepancies could be attributed to the different approaches taken; these latter authors performed stacking on mass-selected galaxies and used a nonevolving FIR–radio
correlation established by \citet{Bell_2003}. The measurements of \citet{2023MNRAS.523.6082C} are systematically higher than ours.
We find that a vertical shift of our Model B SFRD curve will match the \citet{2023MNRAS.523.6082C} data points over the whole redshift range. This is equivalent to multiplying our SFRD by $\sim1.5$, shown as the orange dashed curve. Different from the $q_{\mathrm{IR}}(z)$-based $L_{\mathrm{radio}}-\mathrm{SFR}$ calibration used in the present work, \citet{2023MNRAS.523.6082C} used the calibrated relation between 150 MHz radio luminosity and SFR from \citet{2021A&A...648A...6S}, and also constrained and corrected the scatter in the $L_{150~\mathrm{MHz}}-\mathrm{SFR}$ relation. This may explain the discrepancy between their measurements and ours. As noted by \citet{2020ApJ...899...58L}, the impact of different $L_{\mathrm{radio}}-\mathrm{SFR}$ calibrations is significant.

\subsection{Density evolution is indispensable}
According to \citet{yuan2016mixture}, the evolution of a LF may be regarded as a vector $\vec{E}$, and can be written as
\begin{eqnarray}
        \label{vector}
        \vec{E}=e_1\vec{E}_d+e_2\vec{E}_l,
\end{eqnarray}
where $\vec{E}_d$ and $\vec{E}_l$ are the base vectors of DE and LE, respectively; and $e_1$ and $e_2$ are DE and LE functions as mentioned in Equation (\ref{aaa}). DE carries a physical meaning, and can tell us whether the sources are more or less numerous than those of today, while the LE can tell us whether the sources are systematically more or less luminous than those of today. In all three of our models, the LE function has a peak, indicating that SFGs are, on average, most luminous in radio at z$\sim3-4$. The DE function, according to our LADE models, monotonically decreases with redshift, implying that, in the radio view, SFGs are less numerous in earlier epochs.
From Equation (\ref{eq:SFRD}), we speculate that the shape of the SFRD curve is jointly determined by the form of DE and LE. Comparing Figures \ref{fig:evolution} and \ref{fig:SFRD}, we note that below z$\sim 2$, for all three of our models, the effect of positive LE is dominant over the DE. Therefore, the SFRD curves display a monotonically increasing trend. Above z$\sim 2$, the LE gradually begins to turn over, and the effect of DE begins to show up. As Model C has the strongest negative DE, its SFRD falls more rapidly out to high redshift than Models B and A. One of the main findings of this work is that a DE is genuinely indispensable in modeling the evolution of SFG radio LFs. This finding is further confirmed by the picture that, the assumption of pure LE seems to over-predict the SFRD at high z, while the inclusion of DE corrects for this.

\subsection{The effect of IR--radio correlation}
As shown in Equation (\ref{eq:SFRD}), the calculation of SFRD relies on two components: the LF and the derived SFR. Although our LF has been well constrained (see Section \ref{results}), the calibration of SFR using different scaling factors may also affect the final result. In this section, we show that the SFRD  derived from our model B LF can change depending on the choice of $q_{\mathrm{IR}}(z)$.
The first $q_{\mathrm{IR}}(z)$ we tested is that from \citet{Bell_2003}, where a constant $q_{\mathrm{IR}}$ value of 2.64 was assumed. In Figure \ref{fig:SFRD} (d), the derived SFRD is shown as the gray solid curve. The model seems to significantly over-predict the SFRD above z$ \sim 1.5$.
The second $q_{\mathrm{IR}}(z)$ tested is that from \cite{magnelli2015far}, where $q_{\mathrm{FIR}}(z)=2.35\times(1+z)^{-0.12}$; this relation can be scaled as $\mathrm{log}(L_{\rm FIR})=\mathrm{log}(L_{\rm IR})-\mathrm{log}(2)$ to obtain the $q_{\mathrm{IR}}(z)$.
The SFRD based on this $q_{\mathrm{IR}}(z)$ is presented in the same panel as the gray dashed line, which is generally consistent with our result.
Finally, we considered $q_{\mathrm{IR}}(z)=2.20$ from \cite{2020ApJ...903..138A}. The resultant SFRD is shown as the gray dash-dotted line, and is consistent with our result at $z>3.5$, but is  significantly lower at $z<3.5$. In conclusion, we show that the assumed IR--radio correlation has a significant impact on the derived SFRD and is crucial to accurately constrain the $q_{\mathrm{IR}}$ value at all observed redshifts \citep[also see][]{2017A&A...602A...5N}.

\section{Summary and Conclusions}
\label{summary and comclusions}
In this work, we make use of a star-forming galaxy (SFG) sample (5900 sources) from the VLA-COSMOS 3 GHz data \citep{2017A&A...602A...1S} to measure the radio luminosity functions (LFs). For the first time, we use both models of pure luminosity evolution (PLE) and joint luminosity+density evolution (LADE) to fit the LFs directly to the radio data using a full maximum-likelihood analysis. We fully considered the effect of completeness correction for the sample. We also incorporate the updated observations on local radio LFs and source counts into the fitting process to obtain additional constraints. The parameters of fitting are determined through the Bayesian Markov Chain Monte Carlo (MCMC) approach. In addition, to provide an alternative nonparametric LF estimate for SFGs, we applied the kernel density estimation (KDE) method described in our previous work \citep{yuan2020flexible, yuan2022flexible}. Based on these radio LFs, we derived the dust-unbiased star formation rate density (SFRD) up to a redshift of $z\sim5$. The main results of our work are as follows.

\begin{enumerate}
    \item
Our radio LFs are fitted using three models, assuming a modified-Schechter function with PLE (Model A) and LADE (Models B and C), respectively.
Below $z<2$, the PLE model can fit the radio LFs well, while at higher redshift it gradually deviates from the observations, indicating that PLE is not applicable to describe the evolution at higher redshift. Our LADE models can successfully fit these observed radio LFs from previous studies \citep{Gruppioni_2013,2017A&A...602A...5N,2022ApJ...941...10V} over the whole redshift range, and well reproduce the latest radio source counts from \cite{2020ApJ...903..139A} and \cite{2023MNRAS.520.2668H}. The Akaike information criterion (AIC) also demonstrates that the LADE model is superior to the PLE model.

\item
We find that the luminosity and density evolutions of our LADE models are  broadly consistent with the $L_{\star}$ and $\phi_{\star}$ evolutions inferred by \citet{2023MNRAS.523.6082C} within the uncertainty due to degeneracy of  LE and DE parameters. As the inference of \citet{2023MNRAS.523.6082C}  is model independent, the fact that their results are consistent with ours lends strong support to the efficacy of our LADE models, and we conclude that density evolution is genuinely indispensable in modeling the evolution of SFG radio LFs.

\item
The SFRD curves derived from our PLE and LADE models show a good fit to the SFRD points derived by \citet{2017A&A...602A...5N} and \citet{2022ApJ...941...10V}, respectively. Both their analytical LFs were obtained by fitting the $1/V_{\rm max}$ LF points in all redshift bins, and possibly suffer from bias in the $1/V_{\rm max}$ estimate itself. Our analytical LFs are independent of the $1/V_{\rm max}$ estimates and should be more accurate, which would make our SFRD results an improvement on these previous estimates.

\item
Below $z \sim 1.5$, the SFRD from both our PLE and LADE models is a good match to the multiwavelength compilation from \citet{Madau_2014}, if considering the calibration uncertainty of the FIR–radio relation. Their SFRD curve shows a very similar gradient to ours. This provides a good consistency check for radio emission as a tracer of SFR at low redshift.
Comparing with \citet{Madau_2014}, the SFRD predicted by our three models turns over at a slightly higher redshift ($2<z<2.5$) and falls more rapidly out to high redshift. This was also observed for recent radio estimates by \citet{2022ApJ...941...10V}.

\item
The very recent measurements from \citet{2023MNRAS.523.6082C} are systematically higher than ours (by up to 0.4 dex). This discrepancy could be due to the different $L_{\text{radio}}-\text{SFR}$ calibration and its scatter correction in the work of these latter authors. From the comparison with \citet{2023MNRAS.523.6082C}, we highlight that the $L_{\text{radio}}-\text{SFR}$ calibration and its scatter, or choice of $q_{\rm IR}$ and its evolution, has a significant impact and remains one of the biggest uncertainties in the radio estimates of SFRD.
\end{enumerate}

\begin{acknowledgements}
We thank the anonymous reviewer for the many constructive comments and suggestions, leading to a clearer description of these results. This work is supported by the National Key R\&D Program of China (2023YFE0101200). We acknowledge the financial support from the National Natural Science Foundation of China (grants No. 12073069, No. 12075084, and No.12393813). Z.Y. is supported by the Xiaoxiang Scholars Programme of Hunan Normal University. J.M is supported by the China Manned Space Project (CMS-CSST2021-A06), and the Yunnan Revitalization Talent Support Program (YunLing Scholar project). We would like to thank Yaqian Yu and Yang Liu for helpful discussions.
\end{acknowledgements}

%
%


\appendix

\section{Kernel density estimation}
\label{KDEform}

\begin{figure}
        \centering
        \includegraphics[width=\columnwidth]{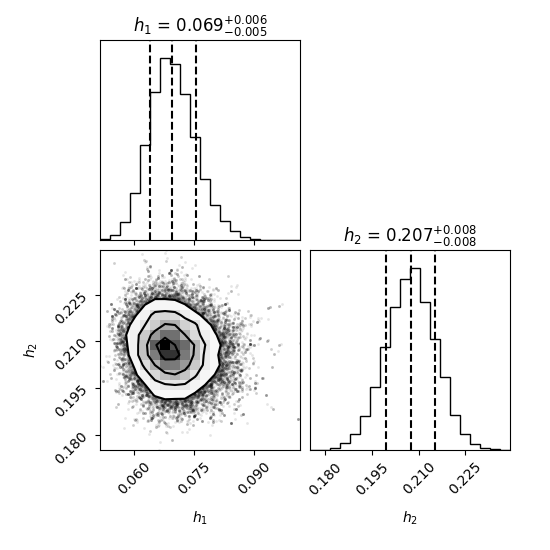}
        \caption{
                Corner plot showing the posterior probability distributions of the bandwidths ($h_1$, $h_2$) of the KDE LF, which were obtained with the routine provided by \citet{yuan2022flexible}. Uncertainties correspond to the 16th and 84th percentiles, while the best-fit value is the median.
        }
        \label{fig:cornerplotkde}
\end{figure}

In many previous studies, the nonparametric LFs are usually derived using the traditional $1/V_{\rm max}$  method of \citet{1968ApJ...151..393S}.
The mathematics behind $1/V_{\rm max}$  is the histogram, the simplest density estimator used mainly for rapid visualization of results in one or
two dimensions. The $1/V_{\rm max}$ method requires the binning of data, which undoubtedly leads to information loss, and potential biases can be caused by evolution within the bins \citep{yuan2020flexible}. In addition, its result is significantly dependent on the choice of bin center and bin width, but currently there are no effective rules to realize an ``optimal'' binning \citep{2013Ap&SS.345..305Y}. To overcome issues surrounding the binning of LFs, \cite{yuan2020flexible, yuan2022flexible} recently proposed a new method for estimating LFs in the framework of kernel density estimation (KDE), the most popular nonparametric density estimation approach developed in modern statistics. This method does not require the binning of data or any model assumptions,
and simultaneously has some advantages and parametric and nonparametric methods. In this work, we apply the KDE method to derive the nonparametric LFs of SFGs.
According to \cite{yuan2020flexible}, the KDE LF can be estimated using
\begin{eqnarray}
\label{trandrf8}
\begin{aligned}
\hat{\phi}(z, L)=\frac{n\left(Z_{2}-Z_{1}\right) \hat{f}\left(x, y \mid h_{1}, h_{2}\right)}{\left(z-Z_{1}\right)\left(Z_{2}-z\right) \Omega \frac{d V}{d z}},
\end{aligned}
\end{eqnarray}
where [$Z_1,Z_2$] is the redshift range of the sample, $n$ is the number of objects in the redshift range, $\Omega$ is the solid angle subtended by the sample, $dV/dz$ is the differential comoving volume per unit of solid angle, and $\hat{f}(x,y|h_{1},h_{2})$ is the density of $(x,y)$, which corresponds to the $(z,L)$ pair in the KDE parameter space. The calculation of $\hat{f}$ can be seen in Equation (9) in \cite{yuan2022flexible}. Its value depends on two bandwidth parameters, $h_1$ and $h_2$.
We use the Python package ``kdeLF'', a Bayesian MCMC routine provided by \cite{yuan2022flexible}, to determine the posterior distributions of the bandwidth parameters (Figure \ref{fig:cornerplotkde}), and then the uncertainty estimation on the LFs. The derived LFs are shown in the panels of Figure \ref{fig:PLFTimeRoman} as red dashed lines with pink contours. The KDE LFs seem to be in good agreement with the binnd LFs, while with smaller uncertainties.

\end{document}